\documentclass[fleqn,usenatbib]{mnras}

\usepackage{newtxtext,newtxmath}

\usepackage[T1]{fontenc}

\DeclareRobustCommand{\VAN}[3]{#2}
\let\VANthebibliography\thebibliography
\def\thebibliography{\DeclareRobustCommand{\VAN}[3]{##3}\VANthebibliography}

\usepackage{graphicx}	
\usepackage{amsmath}	

\title[2Jy radio AGN in the far-IR]{Deep Herschel observations of the 2Jy sample: assessing the non-thermal and AGN contributions to the far-IR continuum}

\author[D.Dicken]{D. Dicken$^{1}$, C. N. Tadhunter$^{2}$, N. P. H. Nesvadba$^{5}$, E. Bernhard$^{2}$,
 V. K\"{o}nyves$^{6}$, R. Morganti$^{3,4}$, 
 \newauthor C. Ramos Almeida$^{7,8}$, and T. Oosterloo$^{3,4}$
\\
$^{1}$UK Astronomy Technology Centre, Royal Observatory Edinburgh, Blackford Hill, Edinburgh EH9 3HJ, UK\\
$^{2}$University of Sheffield, Hounsfield Road, Sheffield, S3 7RH, UK\\
$^{3}$ASTRON, P.O. Box 2,7990 AA Dwingeloo The Netherlands\\
$^{4}$Kapetyn Astronmical Institute, University of Groningen, Postbuss 800, 9700 AV Groningen, The Netherlands\\
$^{5}$Universit\'e de la C\^ote d'Azur, Observatoire de la C\^ote d'Azur, Laboratoire Lagrange, Bd de l’Observatoire, CS 34229, 06304 Nice cedex 4, France \\
$^{6}$Jeremiah Horrocks Institute, University of Central Lancashire, Preston PR1 2HE, UK \\
$^7$Instituto de Astrofisica de Canarias, Calle Via L\'actea, E-38205 La Laguna, Tenerife, Spain \\
$^8$Departamento de Astrofisica, Universidad de La Laguna, E-38206, La Laguna, Tenerife, Spain }

\date{Accepted XXX. Received YYY; in original form ZZZ}

\pubyear{2022}

\begin{document}
\label{firstpage}
\pagerange{\pageref{firstpage}--\pageref{lastpage}}
\maketitle

\begin{abstract}
\noindent
The far-IR/sub-mm wavelength range contains a wealth of diagnostic information that is important for understanding the role of radio AGN in galaxy evolution. Here we present the results of \emph{Herschel} PACS and SPIRE observations of a complete sample of 46 powerful 2Jy radio AGN at intermediate redshifts ($0.05 < z < 0.7$), which represent the deepest pointed observations of a major sample of radio AGN undertaken by \emph{Herschel}. In order to assess the importance of non-thermal synchrotron emission at far-IR wavelengths, we also present new \emph{APEX} sub-mm and \emph{ALMA} mm data. We find that the overall incidence of non-thermal contamination in the PACS bands ($<$200\,$\mu$m) is in the range 28 -- 43\%; however, this rises to 30 -- 72\% for wavelengths ($> $200\,$\mu$m) sampled by the SPIRE instrument. Non-thermal contamination is strongest in objects with compact CSS/GPS or extended FRI radio morphologies, and in those with type 1 optical spectra. Considering thermal dust emission, we find strong correlations between the 100 and 160\,$\mu$m monochromatic luminosities and AGN power indicators, providing further evidence that radiation from the AGN may be an important heating source for the far-IR emitting dust. Clearly, AGN contamination -- whether by the direct emission from synchrotron-emitting lobes and cores, or via radiative heating of the cool dust -- needs to be carefully considered when using the far-IR continuum to measure the star formation rates in the host galaxies of radio AGN.

\end{abstract}




\section{Introduction}

Active Galactic Nuclei (AGN) are increasingly recognised as a key element in the evolution of galaxies of all types. Not only are they likely to be triggered as a consequence of gas infall as galaxies build up their mass, but there is growing evidence that they also directly affect the evolution of their host galaxies through AGN feedback \citep[e.g.][]{nesvadba10,fabian12,harrison17}. In contrast to samples of radio-quiet AGN in the local Universe (e.g. Seyfert galaxies), radio-loud AGN drive powerful relativistic jets that are capable of projecting the power of the AGN into the gaseous haloes of the host galaxies and galaxy clusters, directly affecting the cooling of the hot IGM/ICM gas \citep{mcnamara07, best06}. This makes them particularly important for understanding the link(s) between galaxy evolution and nuclear activity.

A key question concerns how powerful radio-loud AGN are triggered as their host galaxies evolve. A popular approach to investigating this is to examine the detailed optical morphologies  and environments  of the host galaxies. For example, work on the southern 2Jy sample ($0.05 < z < 0.7$)  has demonstrated that 95\% of nearby powerful radio AGN classified as strong-line radio galaxies (SLRG: $EW_{[OIII]} >10$\AA)\footnote{As a class, SLRG include both
type 1 (radio-loud quasars [Q] and broad-line radio galaxies [BLRG]) and type 2 (narrow-line radio galaxies [NLRG]) radio AGN with high equivalent with [OIII]$\lambda$5007 lines; many, but not all, would also be classified as high excitation radio galaxies (HERGs) -- see discussion \citet{tadhunter16}.}  show tidal features at high surface brightness levels \citep{ramos11a,ramos12}; such sources also tend to be found in relatively low density, group-like environments \citep{ramos13}.  In contrast, weak line radio galaxies (WLRG: $EW_{[OIII]} <10$\AA)
show a lower incidence of tidal features and tend to inhabit richer, cluster-like environments \citep{ramos11a,ramos13}. Together, these results provide evidence that, whereas SLRG are predominantly triggered by galaxy mergers and interactions, WLRG are more likely to be fuelled by gas accreted from the hot X-ray haloes of the host galaxies and clusters. However, the optical imaging studies alone provide only a limited view of the triggering events. For example, in the case of galaxy mergers it is difficult to distinguish the nature of the triggering merger (e.g. whether major or minor) from the detection of tidal features alone. Therefore, other observations are required to gain a full picture.

Key additional indicators of the triggering events include the star formation rates (SFR) and cool ISM contents of the host galaxies, which can both be measured via far-infrared (far-IR) observations. Therefore, the sensitivity and resolution at far-IR wavelengths provided by the \emph{Herschel} observatory have been particularly important in this field, and there have been a several \emph{Herschel}-based studies of both the star formation rates \citep{hardcastle13b, kalfountzou14,drouart14,gurkan15,podigachoski15,westhues16} and cool ISM contents \citep{hardcastle13b,tadhunter14,kalfountzou14,podigachoski15} of the hosts of radio AGN. These have provided evidence that SLRG have higher levels of star formation than WLRG \citep{hardcastle13b}, that radio-loud quasars have higher SFR than radio-quiet quasars \citep{kalfountzou14}, and that a significant fraction of high-redshift radio galaxies ($z>1$) are prodigiously star forming, with SFR comparable to or exceeding those of ultra-luminous infrared galaxies 
\citep[ULIRGs,][]{drouart14,podigachoski15}. However, many of these studies have low far-IR detection rates ($<$50\%), and some are based on stacking analysis of shallow, wide-field \emph{Herschel} data. Therefore, it is not always clear whether the results apply to the general radio AGN population, or are biased towards a sub-population that is unusually bright at far-IR wavelengths.

It is also controversial whether the far-IR continuum truly provides a ``clean'' diagnostic of the SFR. In particular, it is often assumed that  any dust heated by the AGN is situated in the compact circum-nuclear torus, is warm and emits predominantly at mid-IR wavelengths, whereas the dust heated by regions of star formation is cooler and emits mainly in the far-IR. However, our \emph{Spitzer} results for the 2Jy sample challenge these assumptions: not only do we find that the [OIII]$\lambda$5007 emission-line luminosity (a proxy for AGN bolometric luminosity) is strongly correlated with the far-IR 70$\mu$m luminosity \citep{dicken09,dicken10}, but also that some 2Jy radio galaxies with high 70$\mu$m luminosities lack independent evidence for recent star formation activity (RSF), such as the detection of strong PAH emission features at mid-IR wavelengths \citep{dicken12}. This suggests that the AGN may play a significant role in heating the cool, far-IR emitting dust.

Furthermore, we face a particular challenge in using far-IR observations for studies of the host galaxies of radio AGN: the potential contamination by the non-thermal synchrotron emission of the cores, jets and lobes of the powerful radio sources. In \citet{dicken08} we attempted to quantify the degree of non-thermal contamination in the far-IR, based on the simple power-law extrapolation of high-frequency ($\sim$20\,GHz) radio core and lobe fluxes to the wavelengths of our \emph{Spitzer} observations for the 2Jy sample. We found that non-thermal contamination is important at 70$\mu$m in up to 30\% of cases. However, this study did not cover the longer far-IR wavelengths ($\ge100\mu$m), which are crucial for the determining precise star formation rates and cool dust masses. Moreover, the extrapolation from radio to far-IR wavelengths was potentially compromised by the lack of photometric measurements in the intermediate mm to sub-mm wavelength region.

Here we present for the first time \emph{Herschel} Photodetecting Array Camera and Spectrometer (PACS)  and Spectral and Photometric Imaging Receiver (SPIRE) photometric observations for the complete 2Jy sample of nearby radio AGN. This is the only photometric survey of a substantial sample of powerful radio AGN undertaken by \emph{Herschel} that is sufficiently deep (detection rates of 100 and 89\% at 100 and 160$\mu$m respectively) to tackle the issues outlined above. Therefore, it provides an important complement to studies based on shallower Herschel surveys of radio AGN. We use these \emph{Herschel} observations along with new \emph{Atacama Pathfinder Experiment} (\emph{APEX}) 870$\mu$m and \emph{Atacama Large Millimeter/sub-Millimeter Array} (\emph{ALMA}) 100\,GHz measurements -- which fill an important gap in the mm to sub-mm SEDs of the sources -- to assess the incidence of non-thermal contamination at far-IR wavelengths. We also revisit the issue of the heating mechanism for the cool dust by investigating whether the 100 and 160$\mu$m luminosities are correlated with AGN power indicators. In a companion paper \citep{bernhard22}, we use these results to determine SFR and cool ISM masses for the 2Jy sample, and compare with other samples of luminous AGN and quiescent non-AGN galaxies measured in a similar way; this allows us to place the triggering of radio AGN in a broader evolutionary context.

The paper is organised as follows. First, in Sections 2 and 3  a discussion of the sample selection is followed by a description of the acquisition and reduction of the \emph{Herschel}, \emph{APEX} and \emph{ALMA} data. Then, in Section 4 we explore the results by plotting the spectral energy distributions (SEDs) for the full 2Jy sample, and assess possible contamination of the thermal infrared emission by non-thermal synchrotron cores, jets and lobes of the radio-loud AGN. Finally, in Section 5 we discuss the origin of the far-IR emission and the evidence that the cool, far-IR emitting dust is heated by radiation from the AGN. 
We assume a cosmology with $H_{o}=71$km s$^{-1} $Mpc$^{-1}, \Omega_{m}=0.27$ and $\Omega_{\lambda}=0.73$ throughout this paper.

\section{Sample selection}
\label{sec:samples}
The 2Jy sample presented in this work is identical to that presented in our previous papers relating to the \emph{Spitzer Space Telescope} observations of powerful southern radio galaxies (\citealp{tadhunter07}; \citealp{dicken08,dicken09,dicken10,dicken12}). It consists of all 46 powerful radio AGN  selected from the 2Jy sample of \citet{wall85} with steep radio spectral indices ($\alpha^{4.8}_{2.7} > 0.5$ for $F_{\nu} \propto \nu^{-\alpha}$), intermediate redshifts (0.05 $<z<$ 0.7), flux densities $S_{2.7\rm{GHz}}>$ 2~Jy, and declinations $\delta<10$\,degrees. 
The spectral index cut has been set to ensure that the radio emission of all the sources in the sample is dominated by steep-spectrum lobe emission, while the lower redshift limit ensures that these galaxies are genuinely powerful radio sources ($P_{5GHz} > 7\times10^{24}$\,W Hz$^{-1}$). For reference, the [OIII]$\lambda$5007 emission-line and 5\,GHz radio luminosities for the sample objects are presented in Table \ref{tab:luminosities} in the Appendix.

The far-IR SEDs analysed in this paper complement a wealth of data that has been obtained for the 2Jy sample over the last two decades. To date, these include: deep optical spectroscopic observations  (\citealp{tadhunter93, tadhunter98, tadhunter02}; \citealp{wills02}; \citealp{holt07}); extensive observations at radio wavelengths (\citealp{morganti93, morganti97, morganti99}; \citealp{dicken08}); complete deep optical imaging from Gemini \citep{ramos11a,ramos12}; deep \emph{Spitzer} MIPS and \emph{Herschel} PACS mid- to far-IR photometric observations (\citealp{dicken08} and this paper: detection rates 100, 90, 100 and  90\% at 24, 70, 100 and 160$\mu$m respectively). In addition, 98\% of the complete sample has been observed at X-ray wavelengths with \emph{XMM} or \emph{Chandra} \citep{mingo14,mingo17}; 93\% have \emph{Spitzer} IRS mid-IR spectra \citep{dicken12,dicken14}; and 78\% have deep 2.2 micron (K-band) near-infrared imaging \citep{inskip10}. 

Our full sample of 46 objects includes a range of optical broad-line radio galaxies and radio-loud quasars (BLRG/Q: 35\%), narrow-line
radio galaxies (NLRG: 43\%)\footnote{Note that the BLRG/Q and NLRG together comprise the SLRG; therefore SLRG make up 78\% of the sample.}, and weak-line radio galaxies  (WLRG: 22\%). In terms of radio morphological classification, the sample
includes 72\% Fanaroff-Riley class II (FRII) sources, 13\% Fanaroff-Riley class I (FRI) sources, and 15\% compact steep spectrum (CSS)/gigahertz peak spectrum (GPS) objects. Detailed information on sample is presented in Tables 1, 2 and A1\footnote{Further, detailed information on individual 2Jy sample objects can be found at http://2jy.extragalactic.info.}. Covering the radio power range $10^{24.8} < L_{5GHz} < 10^{27.9}$\,W Hz$^{-1}$, it is representative of the most powerful radio AGN in the local universe.

\section{Data}
\label{sec:data}

 \begin{table*}
 \begin{minipage}{175mm}
  \caption{Mid-IR to mm-wavelength photometry for the 2Jy sample. For objects observed but not detected, 3$\sigma$ upper limits are given. All fluxes are in mJy, see Section \ref{sec:data} for details.The precise frequencies of the \emph{ALMA} $\sim$100\,GHz ($\sim$3000$\umu$m) observations for individual objects are given in Table \ref{tab:obs_param} in the Appendix.} 
  \label{tab:1}
  \begin{center}
  \begin{tabular}{@{}llrrrrrrrrr@{}}
  \hline 
       &    &{\it Spitzer} & &{\it Herschel} & &{\it Herschel} & & &{\it APEX} &{\it ALMA} \\
       &    &MIPS         & &PACS           & &SPIRE     & & &LABOCA &12m array \\
  \hline
   PKS     &     &  \multicolumn{8}{c}{Flux density (mJy)}\\
   Name & z & 24$\umu$m &  70$\umu$m  & 100$\umu$m  & 160$\umu$m  & 250$\umu$m    & 350$\umu$m  & 500$\umu$m   & 870$\umu$m &$\sim$3000$\umu$m \\
 \hline
0023$-$26	&	0.322 &2.4$\pm$0.3	&26.3$\pm$4.1 	&58.5$\pm$4.3	&86.6$\pm$8.6	&71.7$\pm$5.0	&65.0$\pm$6.9	&30.0$\pm$9.4	&65$\pm$10 & -	\\
0034$-$01	&	0.073 &7.5$\pm$0.2	&17.9$\pm$2.8	&14.7$\pm$3.6	&10.5$\pm$3.6	&	-		&	-		&	-	&	-	&35$\pm$2 \\
0035$-$02	&	0.220	&12.2$\pm$0.1  &23.6$\pm$5.3	&34.4$\pm$4.3	&38.9$\pm$5.3	&68.4$\pm$6.4	&112.7$\pm$7.5	&157.6$\pm$8.6	&$<$225 &186$\pm$9\\
0038$+$09	&	0.188	&25.9$\pm$0.3  &32.2$\pm$5.8	&26.0$\pm$1.7	&23.8$\pm$5.0	&	-		&	-		&	-		&	-	&54$\pm$3		\\
0039$-$44	&	0.346 &33.0$\pm$0.4	&68.7$\pm$8.3	&72.9$\pm$5.6	&61.4$\pm$8.1	&36.8$\pm$4.9	&23.1$\pm$6.5	&$<$22		&$<$15	& -	\\
0043$-$42	&	0.116	&11.1$\pm$0.2  &9.9$\pm$3.2	&15.8$\pm$3.4	&12.9$\pm$3.1	&	-		&	-		&	-		&	-	&4.3$\pm$0.2		\\
0105$-$16	&	0.400	&9.7$\pm$0.2  &$<$11.8		&7.8$\pm$1.5	&$<$7.3  	    &	-		&	-		&	-		&	-	& -		\\
0117$-$15	&	0.565	&6.1$\pm$0.2 &20.2$\pm$2.7	&14.3$\pm$1.5	&$<$9.9     	&	-		&	-		&	-		&	-	& -		\\
0213$-$13	&	0.147	&40.2$\pm$0.1   &37.1$\pm$4.7	&35.2$\pm$3.7	&20.0$\pm$4.1	&	-		&	-		&	-		&	-	&7.4$\pm$0.4	\\
0235$-$19	&	0.620	&11.1$\pm$0.2    &14.3$\pm$2.9\footnote{70{\micron} Flux reported first in \citet{dicken08} now appears uncertain, see notes on objects in the appendix.}	&5.6$\pm$0.8	&$<$4.1	&	-		&	-		&	-		&	-	& -		\\
0252$-$71	&	0.566	&2.9$\pm$0.1 &$<$9.1		&5.9$\pm$1.9	&13.4$\pm$2.8	    &	-		&	-	&	-		&$<$24 & - 	\\
0347$+$05	&	0.339	&3.5$\pm$0.2   &30.8$\pm$5.0	&62.8$\pm$3.3	&84.4$\pm$4.5	&	-		&	-	&	-		&	-	& -		\\
0349$-$27	&	0.066	&8.8$\pm$0.3    &41.9$\pm$4.0	&50.1$\pm$3.8	&46.1$\pm$4.3	&	-		&	-	&	-	&$<$21 &19$\pm$1	\\
0404$+$03	&	0.089	&30.8$\pm$0.1    &70.9$\pm$7.4	&99.3$\pm$5.3	&51.0$\pm$9.2	&	note \footnote{Bright galactic background prevent reliable SPIRE measurements for this object. }	& -		& -		&	-	&18$\pm$1		\\
0409$-$75	&	0.693	&1.5$\pm$0.3 &11.2$\pm$2.0	&45.2$\pm$2.9	&52.0$\pm$6.0	&	-		&	-	&	-		&23$\pm$5 & -	\\
0442$-$28	&	0.147	&22.0$\pm$0.3 &31.0$\pm$5.0	&16.0$\pm$3.8	&$<$11.7		&	-		&	-	&	-		&	-	&29$\pm$2		\\
0620$-$52	&	0.051	&4.5$\pm$0.1    &47.3$\pm$1.4$^{a}$	&17.5$\pm$1.6	&18.1$\pm$4.5	&	-		&	-	&	-		&31$\pm$6 &81$\pm$4	\\
0625$-$35	&	0.055	&24.7$\pm$0.3      &44.8$\pm$5.0	&75.8$\pm$4.0	&88.6$\pm$8.4	&104.0$\pm$7.5	&124.8$\pm$8.4	&149.3$\pm$8.0	&205$\pm$26 &208$\pm$10	\\
0625$-$53	&	0.054	&1.7$\pm$0.2      &$<$10.8	    &9.5$\pm$1.2	&7.1$\pm$2.0	&	-		&	-	&	-		&	-	&18$\pm$1		\\
0806$-$10	&	0.110	&258.3$\pm$0.4      &490.0$\pm$49.0	&516.1$\pm$6.6	&295.0$\pm$8.8	&115.1$\pm$5.6	&36.0$\pm$6.4	&$<$18.6	&	-	&4.8$\pm$0.2	\\
0859$-$25	&	0.305	&9.3$\pm$0.4 &8.4$\pm$2.9	&4.3$\pm$0.9	&$<$9.9		   &	-		&	-	&	-		&	-	& -		\\
0915$-$11	&	0.054	&8.9$\pm$0.2 &110.5$\pm$5.6	\footnote{New 70{\micron} Flux measured from Hershel PACS data. } 	&170.4$\pm$6.6	&149.0$\pm$5.3	&93.8$\pm$4.5	&69.0$\pm$6.8	&80.0$\pm$9.3	&	106$\pm$3 &92$\pm$5	\\
0945$+$07	&	0.086	&47.7$\pm$0.3       &19.4$\pm$4.0$^{a}$	&32.7$\pm$1.9	&30.2$\pm$2.8	&	-	&	-	&	-		&	-	&8.9$\pm$0.05		\\
1136$-$13	&	0.554	&13.8$\pm$0.2       &23.9$\pm$3.7	&29.7$\pm$2.5	&28.9$\pm$3.5	&	-		&	-	&	-		&83$\pm$5	& -		\\
1151$-$34	&	0.258	&16.4$\pm$0.3 &51.5$\pm$6.0	&47.8$\pm$2.0	&51.2$\pm$3.8	&31.8$\pm$5.8	&24.2$\pm$7.4	&36.2$\pm$7.3	&58$\pm$5 &194$\pm$10	\\
1306$-$09	&	0.464	&4.6$\pm$0.2 &21.7$\pm$3.0$^{a}$	&21.5$\pm$3.5	&29.7$\pm$5.0	&	-	&	-	&	-		&84$\pm$5 & -	\\
1355$-$41	&	0.313	&53.1$\pm$0.3   &66.3$\pm$6.9	&72.2$\pm$5.3	&57.2$\pm$4.2	&43.7$\pm$7.6	&$<$21.9   	&$<$26.7 	&	-	& -		\\
1547$-$79	&	0.483	&7.9$\pm$0.1      &19.2$\pm$3.0	&18.4$\pm$2.5	&22.7$\pm$3.2	&	-		&	-	&	-		&	-	& -		\\
1559$+$02	&	0.104	&242.0$\pm$0.3        &470.0$\pm$47.1	&465.5$\pm$5.8	&257.8$\pm$8.4	&100.0$\pm$5.7	&44.4$\pm$8.1	&$<$21.5	&	-	&6.5$\pm$0.3		\\
1602$+$01	&	0.462	&7.7$\pm$0.3          &12.3$\pm$2.8	&16.1$\pm$1.6	&11.6$\pm$2.4	&	-		&	-	&	-		&22$\pm$4	& -		\\
1648$+$05	&	0.154	&2.0$\pm$0.2 &$<$18.8		&16.1$\pm$1.4	&22.6$\pm$2.8	&$<$17.1	&$<$15.9	&$<$29.1 &	-	&0.95$\pm$0.05	\\
1733$-$56	&	0.098	&29.2$\pm$0.3 &151.0$\pm$15.0	&286.8$\pm$7.2	&292.5$\pm$6.7	&155.1$\pm$5.6	&67.0$\pm$8.0	&34.3$\pm$5.5	&38$\pm$12 &59$\pm$3	\\
1814$-$63	&	0.063	&60.6$\pm$0.4      &142.1$\pm$14.0	&186.3$\pm$5.2	&189.0$\pm$7.6	&135.1$\pm$7.7	&128.3$\pm$9.4	&140.8$\pm$12.6	&193$\pm$10 &339$\pm$17	\\
1839$-$48	&	0.112	&3.1$\pm$0.3      &10.9$\pm$3.0	&12.1$\pm$2.1	&37.8$\pm$3.3	&	-		&	-	&	-		&53$\pm$5 &53$\pm$3	\\
1932$-$46	&	0.231	&2.5$\pm$0.1     &17.6$\pm$2.0	&25.0$\pm$3.4	&30.5$\pm$3.4	&	-	    &	-	&	-	    &30$\pm$4 &20$\pm$1	\\
1934$-$63	&	0.183	&17.4$\pm$0.1         &19.9$\pm$2.2$^{a}$     &24.5$\pm$1.3	&35.4$\pm$2.9	&	-		&	-	&	-		&110$\pm$10	\footnote{This is the 95GHz flux from \citet{partridge16}.} &131$\pm$7	\\
1938$-$15	&	0.452	&6.8$\pm$0.3       &19.7$\pm$4.6	&30.5$\pm$1.6	&22.3$\pm$5.4	&	-		&	-	&	-		&	-	& -		\\
1949$+$02	&	0.059	&193.0$\pm$0.2      &348.4$\pm$35.0	&454.2$\pm$6.5	&330.8$\pm$5.2	&148.7$\pm$7.7	&68.3$\pm$8.7	&$<$32.7	&	-	&6.6$\pm$0.3		\\
1954$-$55	&	0.060	&2.7$\pm$0.2 &8.8$\pm$3.0	&9.7$\pm$1.5	&10.8$\pm$3.0	&	-		&	-	&	-		&	-	&29$\pm$2		\\
2135$-$14	&	0.200	&104.9$\pm$0.2 &113.7$\pm$4.8$^{a}$ &140.0$\pm$4.9	&78.6$\pm$7.0	&40.3$\pm$6.7	&19.1$\pm$6.2	&$<$21.9	&20$\pm$4	&56$\pm$3		\\
2135$-$20	&	0.635	&4.3$\pm$0.3 &37.0$\pm$5.7	&80.8$\pm$5.0	&134.3$\pm$6.8	&91.3$\pm$4.4	&55.0$\pm$6.4	&$<$25.1	&43$\pm$11	& -	\\
2211$-$17	&	0.153	&0.5$\pm$0.1 &$<$9.4    	&11.8$\pm$2.3	&24.7$\pm$2.2	&	-		&	-	&	-   	&	-	&1.10$\pm$0.05		\\
2221$-$02	&	0.057	&232.1$\pm$0.3   &186.0$\pm$19.3	&158.9$\pm$4.6	&83.8$\pm$6.7	&$<$24.3	&$<$26.7 &$<$27.3	&	-	&13$\pm$1		\\
2250$-$41	&	0.310	&11.6$\pm$0.1 &22.0$\pm$3.3	&31.9$\pm$1.0	&27.0$\pm$3.4	&	-		&	-	&	-		&	-	& -		\\
2314$+$03	&	0.220	&49.9$\pm$0.3 &512.0$\pm$51.3	&812.4$\pm$4.9	&605.4$\pm$7.3	&285.9$\pm$9.1 &118.5$\pm$8.7	   &50.3$\pm$10.0		   &	-	&34$\pm$2\footnote{The \emph{ALMA} core flux is likely to be contaminated by emission from the nearby eastern radio lobe in this source.} 		\\
2356$-$61	&	0.096	&41.0$\pm$0.2 &74.7$\pm$7.8	&81.0$\pm$2.5	&51.9$\pm$5.5	&32.2$\pm$10.7	    	&$<$39.0   &28.4$\pm$9.3	   &46$\pm$6 &63$\pm$3		\\

\end{tabular}
\end{center}
\end{minipage}
 \end{table*}

Of the 46 objects in the southern 2Jy sample defined in this paper 44 were observed at 100 and 160\micron\ with ESA's \emph{Herschel} \citep{pilbratt10} PACS instrument \citep{poglitsch10} as part of a programme of "Must Do" observations towards the end of the \emph{Herschel} mission (Program ID: \rm{DDT\_mustdo\_4}). The remaining two objects in the sample were observed in separate programs: PKS0915-11 in program \rm{KPOT\_aedge\_1}; and PKS2314+03 in program \rm{OT1\_pogle01\_1}. In addition, the 17 brightest objects at far-IR wavelengths, as quantified by our \emph{Spitzer} observations, were imaged as part of the program DDT\_mustdo\_4 with the \emph{Herschel} SPIRE instrument \citep{griffin10} at 250, 350 and 500\micron\ in order to characterise the shape of their longer-wavelength SEDs, and two further objects were observed with SPIRE as part of the other programs mentioned above. Therefore, 19 objects of the 46 defined in our sample were observed with SPIRE. Objects that are fainter at far-IR wavelengths, as determined using \emph{Spitzer} MIPS observations, were deemed insufficiently bright to detect with SPIRE. Overall, these observations are the deepest pointed observations of a major radio galaxy sample taken by the \emph{Herschel} Observatory, with integration times ranging from 220 to 1570 seconds for the PACS observations, and 721 seconds integration times for the SPIRE observations. 

For the purposes of assessing the degree of possible non-thermal contamination of the mid-infrared photometry, we undertook a program of sub-mm observations of the sample with \emph{APEX} at 870\micron\  using the Large Apex BOlometer CAmera (LABOCA). Of the 46 objects in the sample 33 were observed with \emph{APEX} LABOCA over 3 cycles in periods 88, 89 and 91 from 2011 to 2013 (Program IDs: 088.B-0947(A), 089.B-0791(A) and 091.B-0217(A)). This subset of 33 sources comprises all the objects for which our SEDs and radio maps suggested the possibility of a significant non-thermal contribution to the mid- to far-IR continuum within the \emph{APEX} LABOCA or \emph{Spitzer} beam (based on simple power-law extrapolations from Dicken et al. 2008). Furthermore, all 29 of the sources in our sample with redshifts $z<0.3$ have deep \emph{ALMA} continuum observations at $\sim$100\,GHz, which were taken as part of projects to investigate their CO(1-0) molecular gas properties.  Details of the new \emph{Herschel}, \emph{ALMA} and \emph{APEX} observations are shown in Table \ref{tab:obs_param}.  

In addition to the new photometric results, for completeness  we also present 24 and 70\micron\ fluxes from \citet{dicken08} in Table \ref{tab:1}. Note that, comparison with the new \emph{Herschel} PACS at 100 and 160\micron\ photometry revealed that some of the  70\micron\  \emph{Spitzer} MIPS fluxes are unreliable. In these cases, we obtained an unphysical result when comparing the ratio of the \emph{Spitzer} 70\micron\  fluxes to the 100\micron\ fluxes based on superior  \emph{Herschel} data, and subsequently found evidence for image artifacts and/or confusion that affect the lower S/N and resolution \emph{Spitzer} observations. Such objects are labelled in Table \ref{tab:1} and are detailed in the object descriptions in the Appendix. PKS0915-11 was observed in a different \emph{Herschel} program that also included imaging at 70\micron, therefore we report the updated \emph{Herschel} PACS flux here that benefits from the improved sensitivity and resolution of the \emph{Herschel} observatory at these wavelengths.

\subsection{Herschel data}

The \emph{Herschel Space Observatory} (\emph{Herschel} for short) was a far-IR space telescope that operated between 2009 and 2013. The \emph{Herschel} PACS and SPIRE observations of our sample were taken between October 2012 and March 2013 as part of program \rm{DDT\_mustdo\_4} (see Table \ref{tab:obs_param} for the details of the observations). Preliminary results on the dust masses measured from these data were first presented in \citet{tadhunter14}. The fluxes used in that study, for the identical data set, differ from this work due to the version of the \emph{Herschel} reduction software package used for the data processing: whereas the results presented in \citet{tadhunter14} were processed using HIPE10, those used in this work are based on data processed with HIPE14.2.

The differences between HIPE10 and HIPE14 are numerous, so it is difficult to identify a single cause for changes in the data quality between the images; however, it is clear that the new processed images are superior in both the reduction of artefacts and the noise floor achieved in the imaging. The new fluxes are in general higher by a factor of 1.1 to 1.6 (typically 1.2) both for the PACS and SPIRE bands. The new processing and calibration also brought down the uncertainty in the background for brighter objects in the sample. However, despite these changes, the ratios of the PACS 100 to 160\micron\  fluxes that were used in \citet{tadhunter14} remain remarkably similar, indicating that the calibration changes affect both wavelengths in a similar way: the ratios are the same within 10\% for most objects, rising to a difference of 30\% for a few of the ratios calculated for objects close to the detection limit. Therefore, the conclusions of our previous study \citet{tadhunter14}, which relies on this ratio for the dust mass calculation, still stands with the updated calibration of the data presented here. 

\subsubsection{PACS data}
The PACS data were obtained in the mini-scan-map mode recommended for the point sources, which is how we expect the 2Jy sample targets to appear as at these wavelengths.  Note that the exposure times for the observations differ from object to object, and  were set using the \emph{Spitzer} 70\micron\ flux measurements from \citet{dicken08} as a guide, with the goal of achieving adequate S/N detections for all sources. The data were processed using the ScanmapPhotProject script provided with the HIPE14 software which combines the two angled offset images. This script makes use of an iterative high-pass filtering process of the image timelines to remove the 1/f noise by masking sources over a set threshold. Because some of our sources are relatively faint, we set the mask to the position of the primary source, thereby avoiding fainter targets or the wings of their point response functions (PRFs), that are not detected automatically, being filtered out in the image. The output pixel size was set to the default of 1 arcsecond at 100\micron\  and 2 arcseconds at 160\micron; in comparison, the FWHM of the point spread functions (PSFs) are 7.7 and 12 arcsec respectively at the two wavelengths. The PACS calibration used was 72.0. 

Aperture photometry was performed on the PACS maps inside the ScanmapPhotProject script with a fixed aperture of 10 arcseconds radius at both 100 and 160\micron\ for most sources; however, for  some sources that are faint at 160\micron\ ($<$20\,mJy), we used a smaller aperture of 5 arcseconds radius to limit the influence of background noise on those measurements.  We also used smaller apertures in cases where we needed to separate the flux from nearby companions. Aperture corrections were performed with the photApertureCorrectionPointSource task. 

The depths of the PACS observations have allowed $>$3$\sigma$ detections of 100 and 89\% of the sample at 100 and 160\micron\  respectively. Where the HIPE script could not find the target source automatically - due to bright nearby companion objects or because the target was  intrinsically faint - the fluxes were measured manually with the AnularSkyAperturePhotometry task within HIPE and using the aperture corrections from \citet{balog14}. 

The ScanmapPhotProject script also provided the uncertainties on the PACS measurements, where the 1-sigma uncertainties were estimated by placing adjacent blank sky apertures on the image surrounding the target, while rejecting apertures with companion objects. The apertures size used matched the photometry and a total of 48 apertures were used for the estimation. From these apertures the uncertainty was calculated in two ways: either the median absolute deviation or the standard deviation of the blank sky aperture flux measurements, where  the larger value of the two methods was chosen, conservatively, as the uncertainty. 

\subsubsection{SPIRE data}
In \citet{tadhunter14} we presented results from the SPIRE data reprocessed within the HIPE10 environment, however with HIPE14 and the updated calibration we choose to use the default pipeline processed point source and extended maps from the \emph{Herschel} archive for the analysis, because our reduction was equivalent or no better than those products. The latter HIPE version, using SPIRE calibration 14.2, is most notably improved in terms of instrument artefacts compared with the maps we previously used under HIPE 10. 

The photometry on the sources was performed with the sourceExtractorSussextractor task within the HIPE14 environment using the level 2 data products. This task extracts point sources from an  image using SUSSEXtractor. The script was tailored to fit a point response function (PRF) to sources in the region of interest defined as the position of the target source within the SPIRE images. The script queries the SPIRE Calibration Tree to obtain the appropriate values for the beams, colour corrections and aperture corrections (if used) for a given spectral index ($\alpha$) of the continuum SED, which was assumed to have a power-law shape (i.e. $F_{\nu} \propto \nu^{-\alpha}$). The spectral index  was calculated using an iterative process from the 250 and 350\micron\  fluxes themselves, where an initial guess was made from previous flux measurements in older HIPE versions and then replaced with a calculation based on the output of the new HIPE 14 results. 

In total, 19 objects were observed with SPIRE, where 17 were observed in our ``must-do''program and the 2 remaining objects in other programs as detailed above. The 3$\sigma$ detection rate was 16/19 at 250\micron; 15/19 at 350\micron\  and 10/19 at 500\micron. One object, PKS0404+03, could not be measured at SPIRE wavelengths due to bright Galactic cirrus emission in the SPIRE images. 

In order to represent the flux detections and upper limits with meaningful uncertainties, we choose to utilise a background aperture method similar to that used for the PACS data. The choice of using SUSSEXtractor to measure the flux was driven by the many background confusing sources at these wavelengths that can make aperture photometry difficult. Therefore, the PRF fitting method is more desirable. The SPIRE uncertainties were derived by placing 48 apertures to the background where the aperture size was chosen to be equal to the size of the FWHM of the PSF at each of the three wavelengths: 17.6, 23.9, 35.2 arcseconds for 250, 350 and 500\micron\  respectively. As for the PACS data, the uncertainty quoted is either the median absolute deviation or the standard deviation on all the blank sky aperture flux measurements, whichever was greater. 

\subsubsection{Confused sources}
\begin{figure*} 
\begin{center}
\includegraphics [width=175mm]{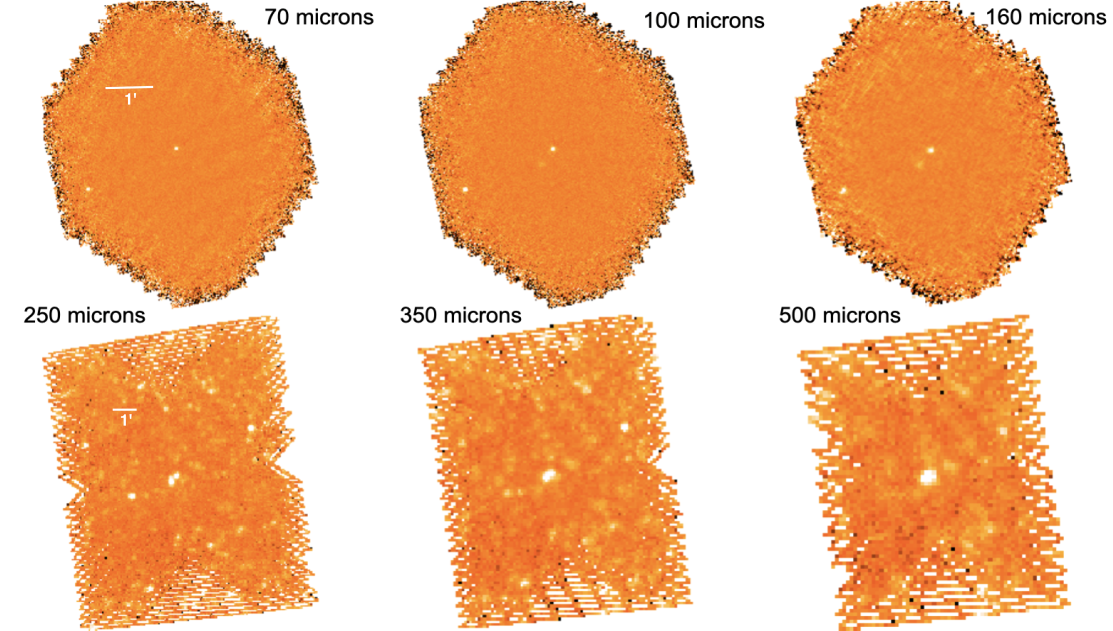}
\caption{\emph{Herschel} PACS (top) and SPIRE (bottom) images for PKS0915-11 and its nearby companion from the \emph{Herschel} archive. The images show how the companion object becomes confused with the source at long wavelengths. Note that the scales are different for the PACS and SPIRE images, and the horizontal bars in the left-hand panels indicate an angular distance of 1\,arcminute on the sky. North is to the top and east to the left. } 
\label{Fig:PKS0915}
\end{center}
\end{figure*}

One of the factors that limited previous far-IR observations of distant AGN was confusion with companion objects in the beam of the instrument. In order to correctly present the SEDs in studies of AGN we should carefully consider contaminating sources which may dominate the flux measured for a point source at longer wavelengths. Gemini imaging of our 2Jy sample \citep{ramos11a,ramos12} enables us to identify the optical counterparts of any sources that may be confused with our targets at mid- and far-IR wavelengths. For many objects, especially those with only PACS data in this study, contamination of the flux measured from nearby companion objects can be avoided by using a smaller aperture to perform the photometry, as described above. However, at longer SPIRE wavelengths the problem of confusion is more apparent. Thanks to the excellent sensitivity and resolution of \emph{Herschel} PACS, we were able to identify potential contaminating sources that may affect the measurements of our SPIRE target images. One object, PKS1934-63, has a nearby companion galaxy that requires us to deblend the fluxes at PACS wavelengths (see below). Of the 19 objects observed with SPIRE, we identify contamination of the measured flux by companion galaxies in 3 targets -- PKS1648+05, PKS2221-02 and PKS2356-61, (17\% of the SPIRE sample) -- at 250\micron, and 4 objects, adding PKS0915-11, (22\% of the SPIRE sample) at 350 and 500\micron. Reduced PACS and SPIRE images of PKS0915-11 can be seen in Figure \ref{Fig:PKS0915}. This object appears brighter than its nearby companion in the PACS bands; however, the companion becomes relatively brighter from 100 to 160\micron. It is clear that the companion then dominates in the longer wavelength SPIRE images.

For the objects that we identified as confused at longer wavelengths we attempted to de-blend the sources. To do this we used GetSources \citep{Men'shchikov12} which is a multi-scale, multi-wavelength source extraction algorithm developed for use in far-IR surveys of star-forming regions with \emph{Herschel}. The method analyses fine spatial decompositions of input images across a range of scales and across all input wavebands. It cleans  single images of noise and background, and constructs wavelength-independent single-scale detection images that preserve information in both spatial and wavelength dimensions. Sources are detected in the combined detection images by following the evolution of their segmentation masks across all spatial scales. Measurements of the source properties were done in the original background-subtracted images at each wavelength; the background was estimated by interpolation under the source footprints and overlapping sources are de-blended in an iterative procedure. We tested the GetSources method using PKS1733-56, an object that did not need de-blending, in order to compare the results. The extracted fluxes agree well with the aperture photometry for PACS and with SUSSEXtractor used for the SPIRE images. Therefore, we were able to use GetSources to extract the fluxes for PKS1934-63 from the PACS images, and the fluxes for PKS0915-11 and PKS2356-61 from the SPIRE images. In Figure \ref{SED-figure3} we plot the PACS fluxes of the source blended with PKS0915-11 so that the reader can get an idea of the potential contamination at SPIRE wavelengths. 

PKS1648+05 and PKS2221-02 are dominated at SPIRE wavelengths by their nearby companion, so we were unable to deblend them even with GetSources. Therefore, we only include upper limits for the fluxes of these object in the SPIRE bands. For PKS2221-02, we have plotted in the SED (Figure \ref{SED-figure6}) PACS measurements of the companion object, so the reader can gauge the degree of contamination to these upper limits. 

\subsection{\emph{APEX}/LABOCA data}

\emph{APEX} is a millimeter/sub-millimeter telescope that comprises a single 12m dish situated on the
Chajnantor plateau in Chile. For this project we used \emph{APEX} with the
LABOCA instrument that operates at a wavelength
of 870$\mu$m.
The data  were taken over three cycles in three programs: from September through November 2011 for ESO Project ID E-088.B-0947A-2011; from May through August 2012 for project ID E-089.B-0791A-2012; and in April 2013 for project ID E-091.B-0217A-2013. See Table \ref{tab:obs_param} for the details of the observations.

For the majority of objects, the observing mode used was chopped photometry, with a symmetric nod i.e., a wobbler was used for the fast switching (i.e. chopping) between the source of interest and an offset position on a selected bolometer. After spending some time with the source in the left beam, the telescope nods to the other side where it continues chopping with the source now in the right beam. The cycle is repeated until the desired depth is reached. Despite being fully commissioned in 2010 there were several problems with the instrument mode that led to our program being re-observed over several observing periods.  This has led to a data set that is not uniform across the sample in its completeness and depth, although the ESO staff did their best to complete our program, taking into account the reduction in expected sensitivity that they encountered. 

To reduce the LABOCA data we used the free astronomical reduction software CRUSH, which has been developed by Attila Kovács and especially designed for use with ground-based millimetre wave cameras.  We used CRUSH version 2.31-1, and followed the advice from the CRUSH LABOCA photometry guide in that we have not reported results from data with only one scan, because it is not easy to determine the reliability in such cases. For multiple scan data we used the reduced-chi-squared value to estimate the significance of the flux detection, where for reduced-chi-squared values larger than unity we scaled the uncertainty by the reduced-chi-squared value. We also removed scans that by eye appeared inconsistent with the rest.  These are likely due to long-timescale systematic effects that degrade the photometric precision, such as pointing drifts, changes in focus quality, and transient presence of emission in the 'OFF' beam as the sky rotates relative to the chopping direction. The estimated uncertainties on the LABOCA photometry include: the high-frequency scatter in the data after correlated components and bad data have been removed; the scatter in the chopped measurements within a nod; and the observed scan to scan scatter. 

PKS0023-26 and PKS2135-20 fluxes were measured from LABOCA maps taken as part of the data run, rather from the wobbler mode data. Aperture photometry was performed on the maps using a 49 arcseconds diameter circular aperture (2.5 beam FWHM), and taking the uncertainty as the  standard deviation of the fluxes measured in eight blank sky apertures.


To complement the LABOCA results we also added flux measurements from the literature:  mm data from \citet{steppe95} and \citet{partridge16} for PKS0023-26 and PKS1934-63 respectively.  

\subsection{ALMA data}

The \emph{ALMA} 12m interferometer operates at millimeter and sub-millimeter wavelengths and comprises an array of at least 43 (and up to 50) 12m dishes situated on the Chajnantor plateau in Chile. All but one of the 29 objects in our sample with redshifts $z < 0.3$ were observed at $\sim$100\,GHz ($\sim$3\,mm) with
\emph{ALMA} in October/November 2019, as part of a project to characterise the molecular C0(1-0) emission in the host galaxies
(Programme: 2019.1.01022.S, PI Tadhunter); the remaining $z < 0.3$
source -- PKS0915-11 -- was observed in October 2016 as part of \emph{ALMA} programme 2017.1.00629.S (PI Edge). The CO observations will be presented in future publications. 
Here we concentrate on the continuum observations that were taken as a by-product 
of the molecular line observations.

The observations were made in 4 baseband windows of frequency width $\sim$1.875\,GHz, one of
which was centred on the expected frequency of the redshifted CO(1-0) line (3840 frequency channels), and the
others on nearby continuum regions (128 frequency channels each). Because of the different redshifts of the
targets, the mean frequency of the continuum windows varies from object-to-object, as reported in Table \ref{tab:obs_param}.
The observations were taken in compact configurations of the 12m array, and this
resulted in beam FWHM in the range 
1.4 -- 3.7\,arcsec. The typical rms sensitivities achieved in the
combined continuum images are in the range 0.01 -- 0.03\,mJy/beam.

Non-thermal core continuum fluxes for the observed 2Jy sources were measured from the standard pipeline-processed average continuum images available from the \emph{ALMA} archive. Cores were detected at $\sim$100\,GHz ($\sim$3\,mm) for all the targets. Since these point-like cores -- with spatial FWHM comparable with those of the ALMA beams
for the observations of individual sources -- were clearly distinguishable from any extended
jet/lobe emission in the maps, we took the peak mJy/beam measurement as the core flux.  
The \emph{ALMA} continuum flux measurements are presented in Table  \ref{tab:1}. The uncertainties in these fluxes are dominated by the accuracy of the absolute flux calibration, which is estimated to be approximately 5\% for the frequency of the observations \citep{warmels18}.

\section{Spectral Energy Distributions}
\label{sec:SED}

\begin{figure*} 
\begin{center}
\includegraphics [width=140mm]{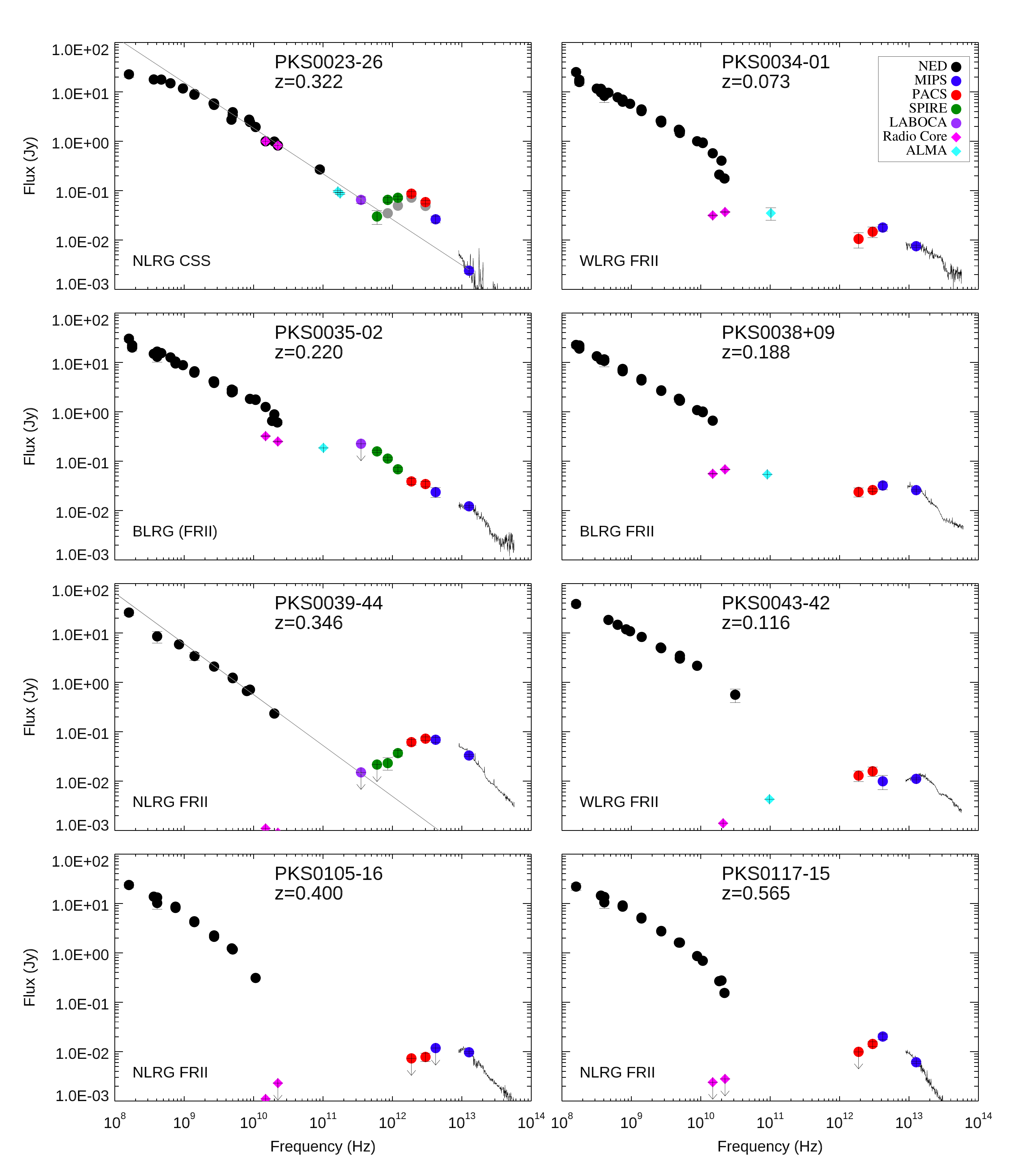}
\caption{Long-wavelength SEDs for 2Jy objects. The black lines at short wavelengths represent the \emph{Spitzer} MIPS spectra, the
grey straight lines show the power-law fits to the high frequency radio to sub-mm data (where relevant), and the grey filled circles the non-thermal
corrected \emph{Herschel} fluxes (only some objects). Note that the
90\,GHz and $\sim$170\,GHz fluxes for PKS0023-26 are taken from \citet{steppe95} and \citet{morganti21} respectively.} 
\label{SED-figure1}
\end{center}
\end{figure*}

\begin{figure*} 
\begin{center}
\includegraphics [width=140mm]{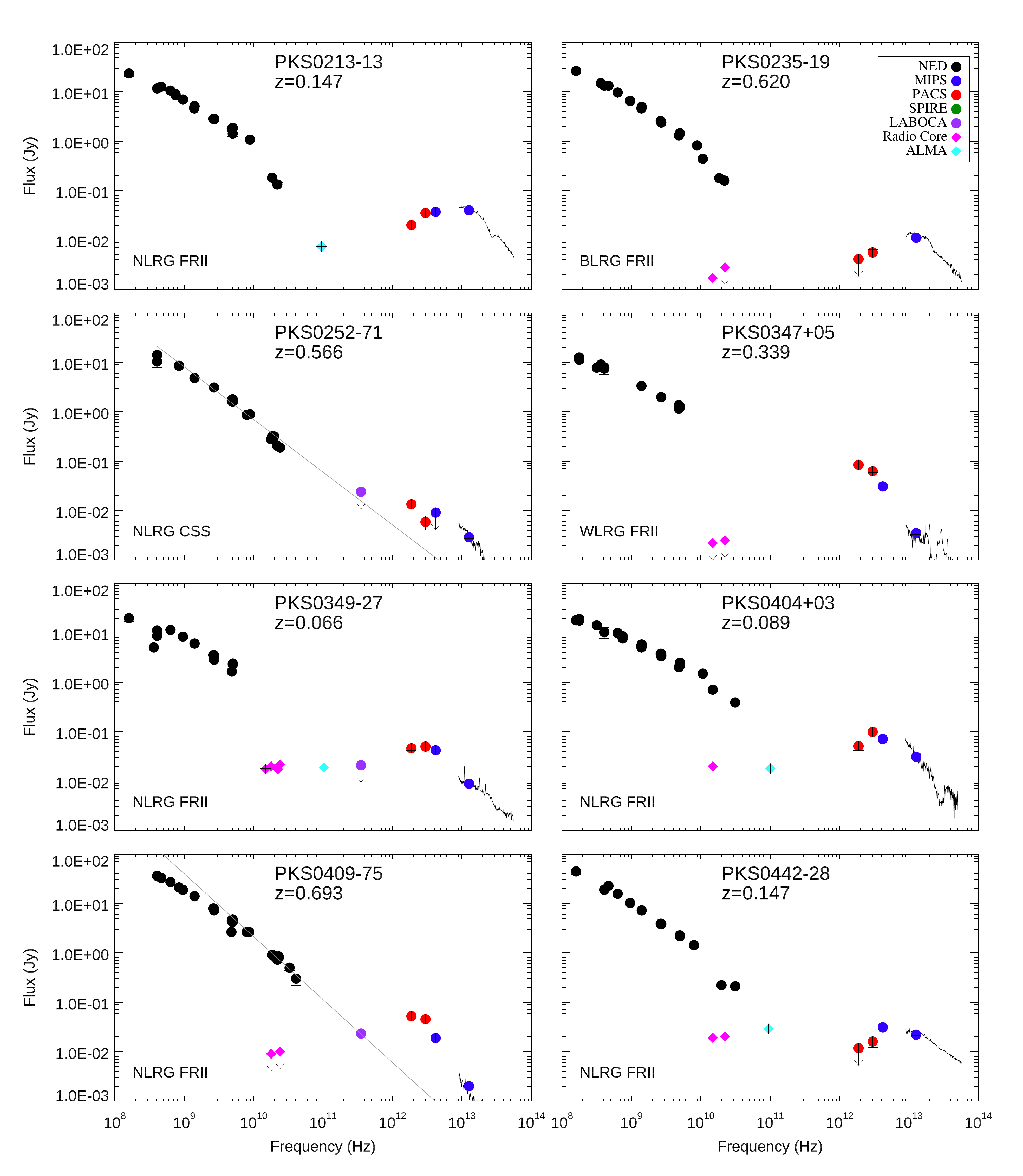}
\caption{Long-wavelength SEDs for 2Jy objects continued (see Figure 2 for details). } 
\label{SED-figure2}
\end{center}
\end{figure*}

\begin{figure*} 
\begin{center}
\includegraphics [width=150mm]{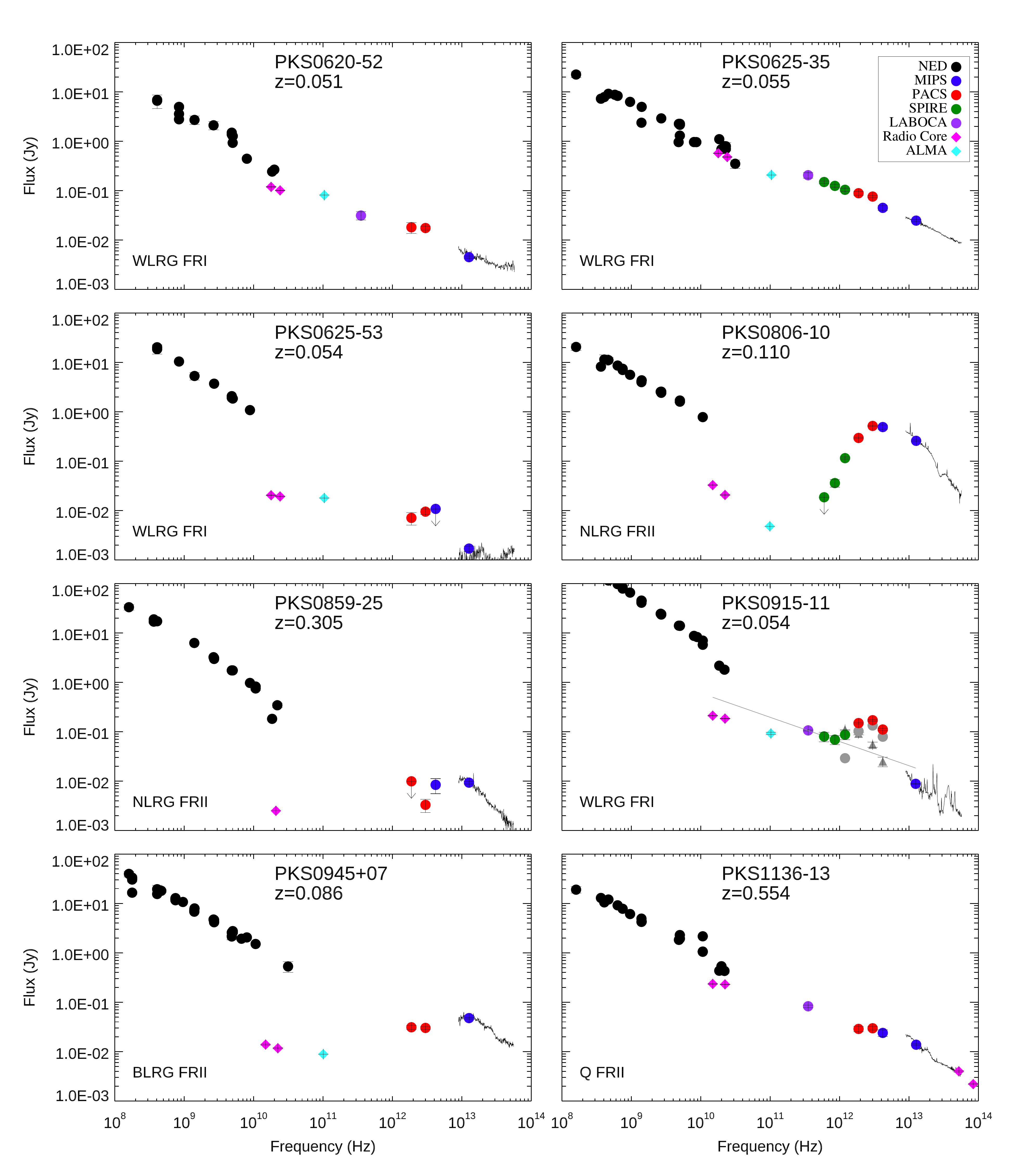}
\caption{Long-wavelength SEDs for 2Jy objects continued (see Figure 2 for details). 
The grey triangles plotted in the SED of PKS0915-16 are for the companion object that is confused with the target at SPIRE bands. Also for this object, the grey line shows the power-law fit to the non-thermal core emission, which was derived using the LABOCA and 500\micron\ SPIRE data points.} 
\label{SED-figure3}
\end{center}
\end{figure*}

\begin{figure*} 
\begin{center}
\includegraphics [width=140mm]{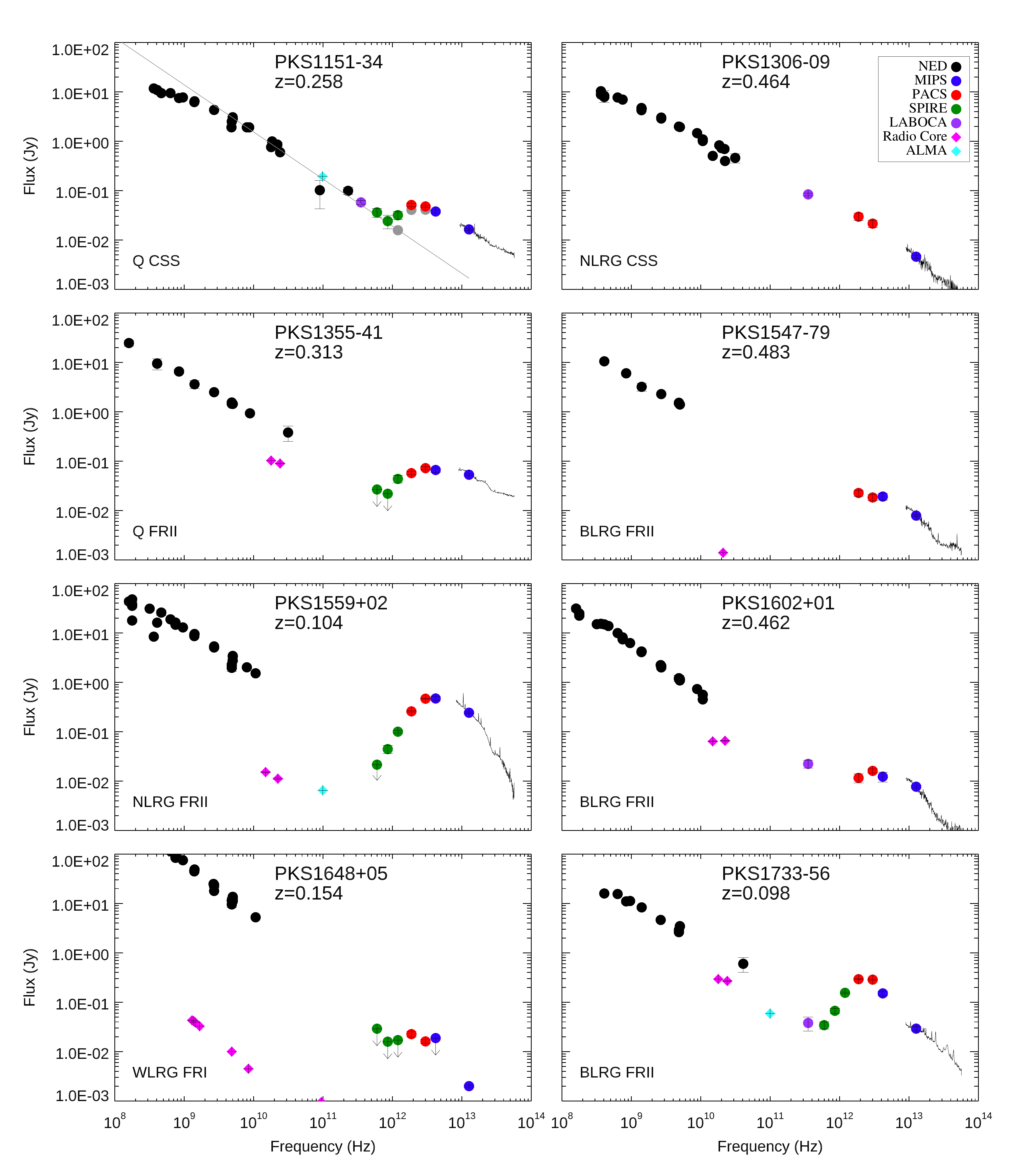}
\caption{Long-wavelength SEDs for 2Jy objects continued (see Figure 2 for details). Note that PKS1648+05 was not observed with \emph{Spitzer} IRS, and the
230\,GHz flux for PKS1151-36 was taken from \citet{steppe95}. } 
\label{SED-figure4}
\end{center}
\end{figure*}

\begin{figure*} 
\begin{center}
\includegraphics [width=140mm]{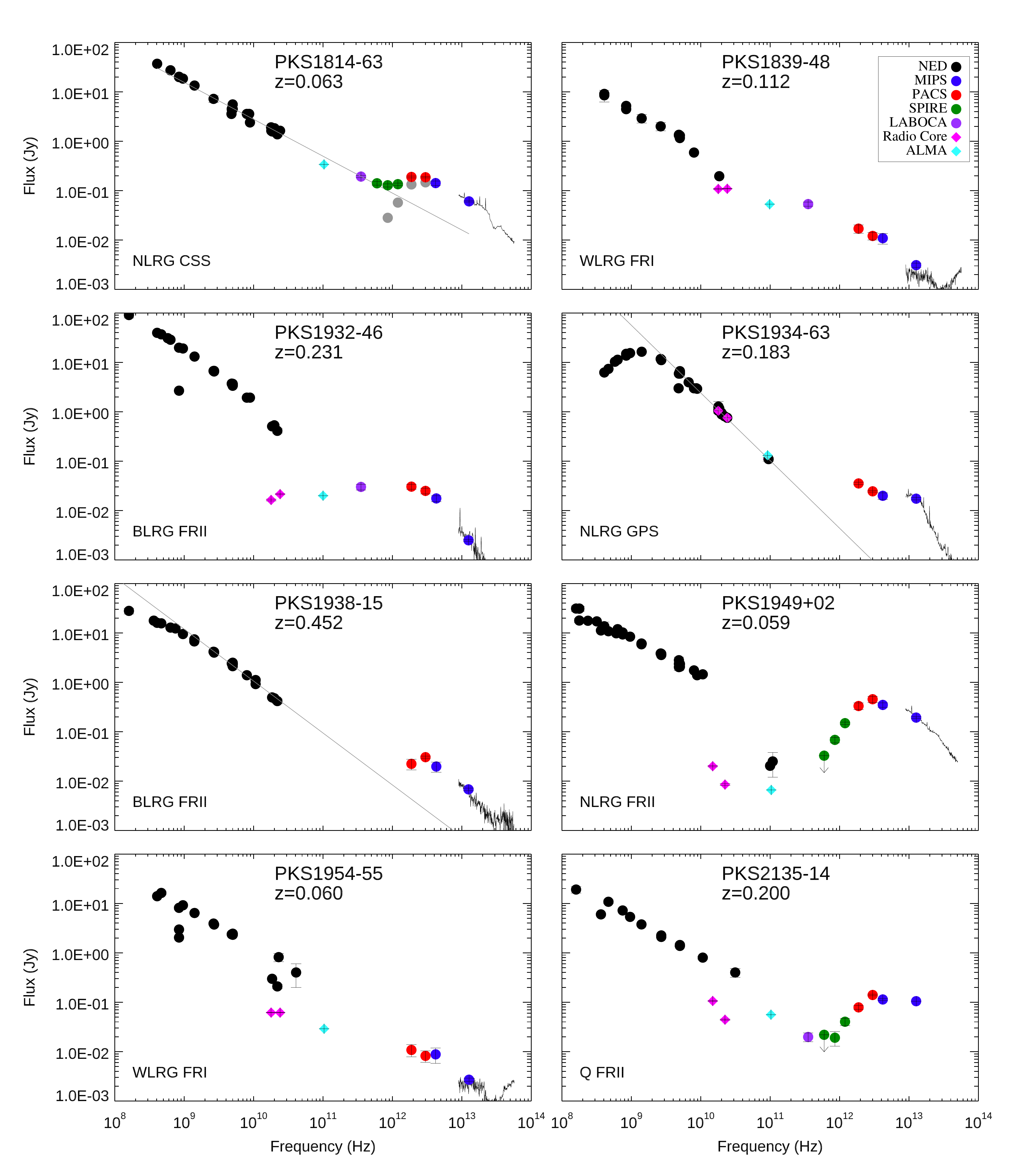}
\caption{Long-wavelength SEDs for 2Jy objects continued (see Figure 2 for details). In addition to the \emph{ALMA} flux measurement, for PKS1934-63 we also plot the 95GHz flux reported in \citet{partridge16}. Lacking sub-mm data, power-laws (grey lines)  have only been fit to the high frequency radio and LABOCA data for PKS1934-63,  and the high frequency radio data alone for PKS1938-15.} 
\label{SED-figure5}
\end{center}
\end{figure*}

\begin{figure*} 
\begin{center}
\includegraphics [width=140mm]{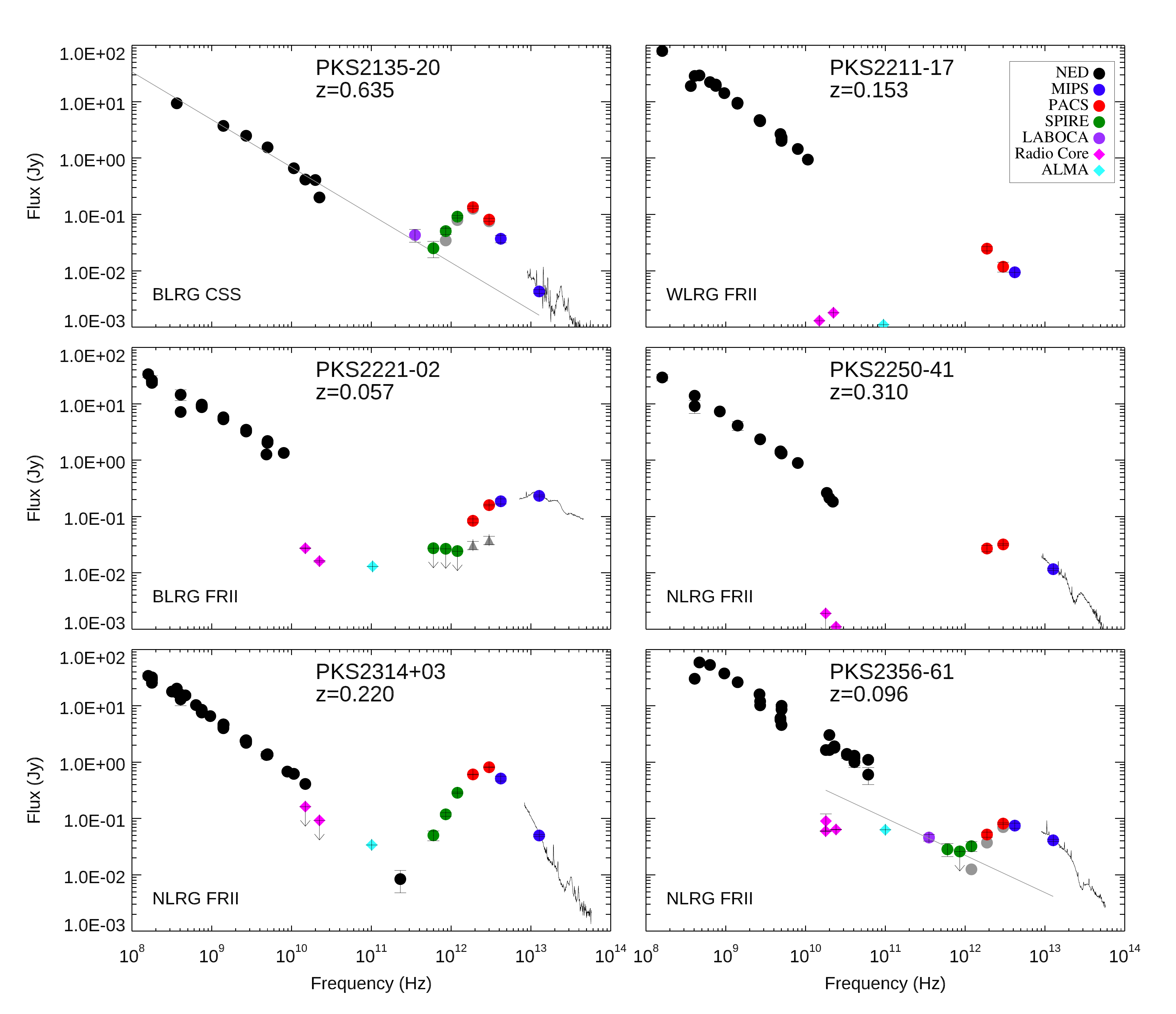}
\caption{Long-wavelength SEDs for 2Jy objects continued (see Fifure 2 for details).   The grey triangles plotted in the SED of PKS2221-02 are for the companion object that is confused with the target at SPIRE bands.  For PKS2356-61 the grey line shows the power-law fit to the non-thermal core emission, which was derived using the LABOCA and 500\micron\ SPIRE data points.} 
\label{SED-figure6}
\end{center}
\end{figure*}

The radio-to-infrared spectral energy distributions (SEDs) of the 46 objects  in our full 2Jy  sample are shown in Figures \ref{SED-figure1} to \ref{SED-figure6}. The new \emph{Herschel} data represent an improvement both in terms of wavelength coverage and image quality (resolution, cosmetics, S/N) on our \emph{Spitzer} programme \citep{dicken08}, which first determined the  SEDs for the 2Jy sample objects, including mid- to far-IR photometry and high frequency radio core data from the \emph{VLA} and \emph{ATCA} observatories. The SEDs now include \emph{Spitzer} IRS spectra from 5 to 30\micron\ presented in \citet{dicken12}, along with the new \emph{Herschel}, \emph{APEX} LABOCA and \emph{ALMA} continuum data presented in this paper. 

The 2Jy SEDs (Figures \ref{SED-figure1} to \ref{SED-figure6}) demonstrate the importance of the new \emph{Herschel}, \emph{APEX} LABOCA and \emph{ALMA} data spanning the wavelength range 100 through 3000\micron\ ($1\times10^{11}$ -- $3\times10^{12}$ Hz), where previously the SEDs suffered a lack of data. The 37\% of the sample with SPIRE data are particularly important, since they help to characterise the shape of the longer-wavelength ($>$200\micron), far-IR continuum SEDs.  With these new data, we are sampling the cool dust components of the powerful radio galaxies, which contain most of the dust mass. 

The diversity in the shapes of the far-IR SEDs of powerful radio AGN is striking. The most prominent features include: objects with strong black-body thermal signatures at far-IR wavelengths (e.g. PKS1559$+$02: Figure \ref{SED-figure4}), objects dominated by non-thermal core emission at both far-IR and sub-mm wavelengths (e.g. PKS0625$-$35: Figure \ref{SED-figure3}), and objects whose SEDs show a thermal bump, but for which there is significant contamination by non-thermal emission from the radio lobes at the longer far-IR wavelengths (e.g. PKS1151$-$34: Figure \ref{SED-figure4}). Despite this apparent diversity, we will show in the next section that many of the differences seen in the SEDs can be explained by varying degrees of non-thermal synchrotron contamination of the thermal dust emission at far-IR wavelengths.

\subsection{Non-thermal contamination}
\label{sec:NT}

We first investigated the possibility of non-thermal contamination of the thermal infrared SED emitted by dust in \citet{dicken08}. Potentially, the relatively high selection frequency of the 2Jy sample ($2.7\rm{GHz}$) could introduce an orientation bias in our results, because at GHz frequencies the radio emission is more likely to contain a component of beamed (flat-spectrum) non-thermal emission than at the lower selection frequencies used for other samples of powerful radio galaxies (e.g. 3CR). Therefore, in order to avoid such an orientation bias, we defined our sample with a steep-spectrum selection ($\alpha \frac{4.8}{2.7} > 0.5$, $F_{\nu} \propto \nu^{-\alpha}$ ), which excludes 16 sources with flatter spectra within the redshift range of the 2Jy sample. However, from examination of the SEDs, it is clear that objects remain in the 2Jy sample that are dominated at far-IR wavelengths by non-thermal emission from the radio cores (see Section 4.1.2 below). Of these objects, only PKS0625-35 has a radio spectral index close to the cutoff for the sample, $\alpha \frac{4.8}{2.7} = 0.53$. Therefore, it is clear that selection on the basis of a steep radio spectrum does not entirely avoid the inclusion in radio AGN samples of some objects that are dominated by non-thermal emission at far-IR wavelengths. There is also the potential for more subtle contamination, in which the non-thermal emission has a significant effect on the far-IR SED shape, while not dominating the flux at these wavelengths. Clearly, we must  consider carefully the potential for non-thermal contamination of the thermal infrared emission.

We consider two types of non-thermal contamination of the thermal infrared emission for the 2Jy sample: {\it lobe/hotspot contamination} due to the steep-spectrum, extended radio lobes and hotspots that fall within the beam of the instrument(s); and {\it radio core contamination} due to the emission of the inner jets which, due to self-absorption effects, often has a flat spectrum at radio wavelengths. We defer discussion the potential effects of 
variability of the non-thermal core emission to the end of
section Section \ref{sec:radio_optical}.


With the \emph{Herschel}, \emph{APEX} and \emph{ALMA} data added to the SEDs, we now investigate the degree of non-thermal contamination of the far-IR dust emission of the 2Jy sample. We present the results of the new analysis in Table \ref{tab:NTcontam}, in which objects are ordered according to their optical and radio classes; we also distinguish between longer ($>$200\micron) and shorter($<$200\micron) far-IR wavelengths, because of the possible wavelength dependence in the contamination. 

For each wavelength range, we mark objects in which the non-thermal emission dominates in a particular wavelength range with a ``D'', those with a potential for non-thermal contamination with a ``P'', and those with non-thermal contamination that can be corrected by a ``C''. Organised in this way, the table reveals clear trends in the degree of non-thermal contamination that are related to the classifications of the sources. 

\subsubsection{Non-thermal contamination by extended jets, lobes or hotspots}

Contamination of the thermal infrared emission by the non-thermal synchrotron radiation of the extended lobes, jets and hotspots is only important for those objects in which the extended radio components fall within the beam of the infrared or sub-mm instrument(s). Moreover, it is more likely to affect the longer infrared wavelengths, first because the beam of the instrument tends to be larger at long wavelengths, and second because the steep-spectrum synchrotron emission is stronger at such wavelengths. Therefore, we investigated such contamination for all the 2Jy objects with extended non-thermal components within the \emph{Herschel} SPIRE diffraction limited beams of FWHM 18, 25 and 36\,arcseconds for the 250, 350 and 500\micron\ bands respectively, although PACS and \emph{Spitzer} data will potentially be affected as well, as we will discuss. To make this assessment we used both  our high frequency \emph{ATCA/VLA}  radio maps and \emph{ALMA} 100GHz continuum images (where  available).

From columns 6 and 7 in Table \ref{tab:NTcontam} it is striking that five out of the seven compact steep-spectrum (CSS) or gigahertz-peak spectrum (GPS)  (column 5) sources in the 2Jy sample are affected by non-thermal contamination from their radio lobes or hotspots at far-IR wavelengths. For all these compact ($D < 20$\,kpc) radio sources the radio lobe/hotspot emission is entirely contained within the \emph{Herschel} beam for all SPIRE and PACS wavelengths; consistent with the results of \citet{murgia99}, the new data demonstrate that such sources do not show
the high frequency spectral steepening that is sometimes observed for the lobes of more extended radio sources \citep{cleary07}.

For many of the objects with compact radio sources we see non-thermal contamination at both shorter ($<$200\micron) and longer ($>$200\micron) far-IR wavelengths; however, for four of the seven CSS/GPS objects (PKS0023-26, PKS1151-34, PKS1814-63 and PKS2135-20) it is possible to correct at least the shorter-wavelength far-IR fluxes for this contamination by using a simple power-law extrapolation of the radio-to-sub-mm SEDs (marked with a ``C''in Table \ref{tab:NTcontam}).  The corrected fluxes are shown in Table \ref{tab:NTcor}, which benefit from the added constraints on the SEDs of the non-thermal component provided by the LABOCA and longer-wavelength SPIRE data. The power-law fits to the radio and sub-mm data used to calculate the results are plotted on the SEDs of each object. 

For two of the remaining three CSS/GPS (PKS0252-71 and PKS1934-63), we find that the non-thermal contamination is likely to be negligible in the PACS bands\footnote{Note that, although the non-thermal contamination is likely to be negligible in PACSs band for these objects, it could be significant at longer far-IR wavelengths, but we lack the SPIRE data to test this.}. However, in the case of the third object (PKS1306-09) non-thermal emission completely dominates the infrared SED, so it is not possible to clearly  identify a thermal dust component. We note that the presence of a dominant non-thermal component in this source is  consistent with its optical polarization properties \cite[see][]{tadhunter94a}.

We have also identified nine additional objects where there is a potential for contamination by extended non-thermal components, but for which this contamination is uncertain. These objects are marked with a ``P'' in Table \ref{tab:NTcontam} columns 6 and 7. Four of them (PKS0035-02, PKS0625-35, PKS1136-13, PKS1839-48) also have strong non-thermal contamination from their cores, as discussed below. In such objects, while the core contamination dominates,  there may also be a contribution from extended jets, lobes or hotspots, based on their radio size and the extrapolation of the extended radio emission through to the infrared part of the SED. Indeed, two of these objects show prominent, one-sided jets in optical images \citep[PKS0625-35, PKS1136-13:][]{ramos11a}, and three (PKS0035-02, PKS0625-35, PKS1839-48) show relatively strong emission from their jet/lobes/hotspots on the scale of the \emph{Herschel} PACS beam  in our \emph{ALMA} $\sim$100GHz images.

For the remaining five objects with uncertain extended jet/lobe/hotspot contribution (PKS0034-01, PKS0039-44, PKS0409-75, PKS1938-15, PKS2250-41), the uncertainty is exclusively for the longer wavelengths $>$200\micron.  Of these, all but PKS0034-01 are at the higher redshift end of the sample ($0.34 < z < 0.7$). Indeed, the NRLG/FRII PKS0409-75 ($z = 0.693$) is both the most distant radio source and the most powerful at radio wavelengths in the 2Jy sample. Although not observed with SPIRE, the LABOCA sub-mm flux falls directly on a power-law extrapolation of the high frequency radio SED for this object. Because of the relatively small angular extent of the source (largest angular size $= 9$\,arcsec or 85 kpc), which is dominated by two lobes of emission \citep{morganti99}, we might expect the radio lobes to contribute at the longer far-IR wavelengths. This emphasises that the powerful, steep-spectrum emission from the lobes of high-z radio galaxies can make a substantial flux contribution at the longer far-IR wavelengths, potentially compromising SFR and dust mass estimates based on fluxes measured at those wavelengths. See further discussion in Section \ref{sec:disc}. 

To summarise, a total of 9/46 objects in our sample (20\%) suffer clear or possible contamination by non-thermal lobe/jet/hotspot emission at shorter far-IR wavelengths (i.e. PACS bands: $<$200\micron). For one of these objects the non-thermal emission dominates in the far-IR to the extent that a thermal dust component cannot be identified, but for four of the remaining sources it is possible to correct the non-thermal contamination, based on a simple power-law extrapolation of the longer-wavelength data. However, a higher proportion of the sample (16/46 or 35\%) show clear or potential lobe/jet/hotspot contamination at longer far-IR wavelengths 
(i.e. SPIRE bands: $>$200\micron); in many cases this contamination cannot
be readily corrected due to the dominance of the non-thermal emission, or a lack of the sub-mm data required for accurate extrapolation.

\subsubsection{Non-thermal core contamination}
\label{sec:NTcore}

Determining the degree of non-thermal contamination by the compact core components is potentially more challenging than it is for the steep-spectrum extended components. This is because a simple power-law extrapolation will not always suffice. For example, a synchrotron 
core source could show a relatively flat spectrum in $F_{\nu}$ from radio to sub-mm wavelengths, but then a sharp turnover to a steeper spectrum in the infrared, as the emitting plasma becomes optically thin \citep[e.g. NGC1052:][]{ontiveros19}. However, for the majority of cases with well-sampled SEDs in the literature and in the 2Jy sample itself, the SEDs of the non-thermal cores are either flat or decline with wavelength in the mm to infrared wavelength range \citep[e.g.][]{dicken08,cleary07,prieto16}. The difficulty in determining the degree of non-thermal contribution from the core components is further exacerbated by the fact that some of the sources for which this form of non-thermal contamination might be important lack the constraints that would be provided by LABOCA or SPIRE data (e.g. PKS0034-01 and PKS0038+09 in Figure 2).

In cases where the core fluxes measured using radio, \emph{ALMA} and LABOCA data fall well above the 
far-IR fluxes, and the far-IR fluxes appear to be an extrapolation of the radio to sub-mm SED of the core, without
any clear evidence for a far-IR thermal bump (i.e. 160\micron\, flux larger than or comparable with the 100\micron\, flux), we consider that the non-thermal core emission is likely to be 
dominant in the far-IR (marked by a ``D'' in Table \ref{tab:NTcontam}). Six objects in our full sample fall into this category:
PKS0035-02, PKS0620-52, PKS0625-35, PKS1136-13,  PKS1839-48, and PKS1954-55.

There are also objects with possible non-thermal core contamination at far-IR wavelengths, but for which there is some uncertainty because of a lack of data at sub-mm wavelengths,  or possible evidence for a thermal bump in the far-IR (i.e. 160\micron\, flux lower than the 100\micron\, flux). To identify these objects,  we adopt the criterion that any object for which the highest frequency
radio to sub-mm core flux (usually the \emph{ALMA} 100GHz core flux but in some cases the \emph{APEX} LABOCA flux)  is $>$50\% the flux at a particular far-IR wavelength has a  potential for non-thermal core contamination, and mark the object with a ``P'' in the table. In this case, we are making the assumption that the SED of the non-thermal core remains flat up to far-IR wavelengths\footnote{A further is assumption is that the \emph{ALMA} observations, despite their relatively low spatial resolution (beam $FWHM\sim$1.4 -- 3.7 arcsec), are dominated by the compact cores that are unresolved in the higher-resolution ($<$1 arcsec) $\sim$20\,GHz \emph{ATCA/VLA} observations, and do not suffer from significant contamination by steeper-spectrum
jet or hotspot emission on the kpc-scales covered by the \emph{ALMA} beam. However, examination of the \emph{ATCA/VLA} maps shows that such contamination is likely to be negligible in most cases, and for the few objects where it may be significant, the overall radio to far-IR SEDs already suggest that the non-thermal core emission dominates. Therefore, this does not affect the statistics on non-thermal core contamination of the far-IR.}. However, this is likely to be conservative, given that the SEDs of most compact core sources with more complete data decline in flux over this wavelength range. Using
this criterion, six objects show  potential for non-thermal core contamination at far-IR wavelengths: PKS0034-01, PKS0038+09, PKS0442-28, PKS0625-53, PKS1602+01 and PKS1932-46. Moreover, PKS0859-25 is
also classed as a P object, based on a flat extrapolation of its 20\,GHz ATCA radio core flux. However, this case is highly uncertain, due to the lack of any mm or sub-mm photometric points. 

In addition, there are two objects -- PKS0915-11 and PKS2356-61 -- for which the longer far-IR wavelengths are clearly dominated by non-thermal core emission,
but whose well-sampled radio-to-far-IR SEDs (including \emph{ALMA}, \emph{APEX} LABOCA and \emph{Herschel} SPIRE data) show evidence for a turnover to steeper spectra at $\sim$350 - 500\micron. In these cases, a simple power-law extrapolation of the LABOCA to longer-wavelength SPIRE points allows us to estimate the contribution
of the non-thermal contamination on the shorter-wavelength fluxes. We mark these objects with a "C" in
Table \ref{tab:NTcontam}, and Table \ref{tab:NTcor} shows the far-IR fluxes for these two sources after this non-thermal core power-law has 
been subtracted.

We note that the non-thermal core contamination we have considered so far will potentially affect all far-IR wavelengths (both PACS and SPIRE bands). However, there are seven additional objects for which  non-thermal core contamination is likely, or has the potential, to be significant in the
SPIRE bands alone, based on the extrapolation of the radio-to-sub-mm SEDs: PKS0043-46, PKS0213-13, PKS0349-27, PKS0404+03, PKS1355-41, PKS1733-56 and PKS2135-14. 

Overall, a maximum of 15/46 of our full sample (33\%) show evidence for contamination by the emission of the non-thermal core sources in the shorter-wavelength PACS bands, and a maximum of 22/46 (48\%) in the longer-wavelength SPIRE bands. However, two cases  have sufficiently well-sampled SEDs to allow estimation of the contribution of the non-thermal core emission to
the PACS fluxes.

 \begin{table*}
 \begin{minipage}{175mm}
  \caption{2Jy sample non-thermal contamination results. The Table is organised by optical class: Broad-line/Quasar (BLRG/Q), narrow-line radio galaxy (NRLG) and weak-line radio galaxy (WLRG); and also sorted by radio class: FRI, FRII and compact steep spectrum (CSS)/gigahertz peak spectrum (GPS) objects. Lobe/hotspot and core contamination are shown separately, as are the results above and below 200\,$\micron$. Objects with non-thermal contamination are identified as follows: $\bf{C}$ indicates objects with significant non-thermal contamination, but where the contamination is correctable for at least some of the photometric bands in the wavelength range - see Table \ref{tab:NTcor}; $\bf{D}$ indicates cases where the SED is dominated by the non-thermal emission that cannot be corrected; and $\bf{P}$ indicates objects with a strong potential for non-thermal contamination, but for which the level of this contamination is uncertain. Column 10 indicates whether the objects has strong, independent evidence for recent star formation activity \citep[RSF -- see][]{dicken12}.}
  \label{tab:NTcontam}
  \begin{center}
  \begin{tabular}{@{}llcccccccc@{}}
  \hline
   (1)  & (2)  & (3) & (4)  &  (5)  & (6)     & (7)              & 
   (8)                   & (9)                  & (10) \\
   Name  & Other  & z & Optical  &  Radio      & Lobe              & Lobe                   & Core                  & Core & Star   \\
   	      &  Name &   & Class  &   Class     &  $<$200\micron & $>$200\micron &  $<$200\micron & $>$200\micron & Formation \\
 \hline
 \hline
 BLRG and Q&&&&&&&\\		
 \hline
1151$-$34	&		&	0.258	&	Q	&	CSS	&	$\bf{C}$	&	$\bf{D}$	&	-	&	-	&	-	\\
2135$-$20	&	OX-258	&	0.635	&	BLRG	&	CSS	& $\bf{C}$	&	$\bf{C}$	&	-	&	-	&	RSF	\\
&&&&&&&\\																			
0035$-$02	&	3C17	&	0.220	&	BLRG	&	(FRII)	&	$\bf{P}$	&	$\bf{P}$	&	$\bf{D}$	&	$\bf{D}$	&	-	\\
0038$+$09	&	3C18	&	0.188	&	BLRG	&	FRII	&	-	&	-	&	$\bf{P}$	&	$\bf{P}$	&	-	\\
0235$-$19	&	OD-159	&	0.620	&	BLRG	&	FRII	&	-	&	-	&	-	&	-	&	-	\\
0945$+$07	&	3C227	&	0.086	&	BLRG	&	FRII	&	-	&	-	&	-	&	-	&	-	\\
1136$-$13	&		&	0.554	&	Q	&	FRII	&	$\bf{P}$	&	$\bf{P}$	&	$\bf{D}$	&	$\bf{D}$	&	-	\\
1355$-$41	&		&	0.313	&	Q	&	FRII	&	-	&	-	& -		&	$\bf{P}$	&	-	\\
1547$-$79	&		&	0.483	&	BLRG	&	FRII	&	-	&	-	&	-	&	-	&	-	\\
1602$+$01	&	3C327.1	&	0.462	&	BLRG	&	FRII	&	-	&	-	&	$\bf{P}$	&	$\bf{P}$	&	-	\\
1733$-$56	&		&	0.098	&	BLRG	&	FRII	&	-	&	-	&	-	&	$\bf{P}$	&	RSF	\\
1932$-$46	&		&	0.231	&	BLRG	&	FRII	&	-	&	-	&	$\bf{P}$	&	$\bf{P}$	&	RSF	\\
1938$-$15	&		&	0.452	&	BLRG	&	FRII	&	-	&	$\bf{P}$	&	-	&	-	&	-	\\
2135$-$14	&		&	0.200	&	Q	&	FRII	&	-	&	-	&	-	&	$\bf{D}$	&	-	\\
2221$-$02	&	3C445	&	0.057	&	BLRG	&	FRII	&	-	&	-	&	-	&	-	&	-	\\
 \hline																			
NLRG &&&&&&&\\								 											
 \hline																			
0023$-$26	&		&	0.322	&	NLRG	&	CSS	&	$\bf{C}$	&	$\bf{C}$	&	-	&	-	&	RSF	\\
0252$-$71	&		&	0.566	&	NLRG	&	CSS	&	-	&	$\bf{P}$	&	-	&	-	&	-	\\
1306$-$09	&		&	0.464	&	NLRG	&	CSS	&	$\bf{D}$	&	$\bf{D}$	&	-	&	-	&	-	\\
1814$-$63	&		&	0.063	&	NLRG	&	CSS	&	$\bf{C}$	&	$\bf{C}$	&	-	&	-	&	-	\\
1934$-$63	&		&	0.183	&	NLRG	&	GPS	&	-	&	$\bf{P}$	&	-	&	-	&	-	\\
&&&&&&&\\																			
0039$-$44	&		&	0.346	&	NLRG	&	FRII	&	-	&	$\bf{P}$	&	-	&	-	&	-	\\
0105$-$16	&	3C32	&	0.400	&	NLRG	&	FRII	&	-	&	-	&	-	&	-	&	-	\\
0117$-$15	&	3C38	&	0.565	&	NLRG	&	FRII	&	-	&	-	&	-	&	-	&	-	\\
0213$-$13	&	3C62	&	0.147	&	NLRG	&	FRII	&	-	&	-	&	-	&	$\bf{P}$	&	-	\\
0349$-$27	&		&	0.066	&	NLRG	&	FRII	&	-	&	-	&	-	&	$\bf{P}$	&	-	\\
0404$+$03	&	3C105	&	0.089	&	NLRG	&	FRII	&	-	&	-	&	-	&	$\bf{P}$	&	-	\\
0409$-$75	&		&	0.693	&	NLRG	&	FRII	&	-	&	$\bf{P}$	&	-	&	-	& RSF		\\
0442$-$28	&		&	0.147	&	NLRG	&	FRII	&	-	&	-	&	$\bf{P}$	&	$\bf{P}$	&	-	\\
0806$-$10	&	3C195	&	0.110	&	NLRG	&	FRII	&	-	&	-	&	-	&	-	&	-	\\
0859$-$25	&		&	0.305	&	NLRG	&	FRII	&	-	&	-	&	$\bf{P}$	&	$\bf{P}$	&	-	\\
1559$+$02	&	3C327	&	0.104	&	NLRG	&	FRII	&	-	&	-	&	-	&	-	&	-	\\
1949$+$02	&	3C403	&	0.059	&	NLRG	&	FRII	&	-	&	-	&	-	&	-	&	-	\\
2250$-$41	&		&	0.310	&	NLRG	&	FRII	&	-	&	$\bf{P}$	&	-	&	-	&	-	\\
2314$+$03	&	3C459	&	0.220	&	NLRG	&	FRII	&	-	&	-	&	-	&	-	&	RSF	\\
2356$-$61	&		&	0.096	&	NLRG	&	FRII	&	-	&	-	&	$\bf{C}$	&	$\bf{D}$	&	-	\\
 \hline																			
WLRG &&&&&&&\\								 											
 \hline																			
0620$-$52	&		&	0.051	&	WLRG	&	FRI	&	-	&	-	&	$\bf{D}$	&	$\bf{D}$	& RSF		\\
0625$-$35	&	OH-342	&	0.055	&	WLRG	&	FRI	&	$\bf{P}$	&	$\bf{P}$	&	$\bf{D}$	&	$\bf{D}$	&	-	\\
0625$-$53	&		&	0.054	&	WLRG	&	FRI	&	-	&	-	&	$\bf{P}$	&	$\bf{P}$	&	-	\\
0915$-$11	&	Hydra A	&	0.054	&	WLRG	&	FRI	&	-	&	-	&	$\bf{C}$	&	$\bf{D}$	&	RSF	\\
1839$-$48	&		&	0.112	&	WLRG	&	FRI	&	$\bf{P}$	&	$\bf{P}$	&	$\bf{D}$	&	$\bf{D}$	&	-	\\
1954$-$55	&		&	0.060	&	WLRG	&	FRI	&	-	&	-	&	$\bf{D}$	&	$\bf{D}$	&	-	\\
&&&&&&&\\																			
1648$+$05	&	Herc A	&	0.154	&	WLRG	&	FRI/FRII	&	-	&	- 	&	-	&	-	&	-	\\
0034$-$01	&	3C015	&	0.073	&	WLRG	&	FRII	&	-	& $\bf{P}$	& $\bf{P}$	&	$\bf{P}$	&	-	\\
0043$-$42	&		&	0.116	&	WLRG	&	FRII	&	-	&	-	&	-	&	$\bf{P}$	&	-	\\
0347$+$05	&		&	0.339	&	WLRG	&	FRII	&	-	&	-	&	-	&	-	&	RSF	\\
2211$-$17	&	3C444	&	0.153	&	WLRG	&	FRII	&	-	&	-	&	-	&	-	&	-	\\													

\end{tabular}
\end{center}
\end{minipage}
 \end{table*}

 \begin{table*}
  \caption{Non-thermal corrected \emph{Herschel} flux results. Due to the uncertainty in the sub-mm point (\emph{ALMA} or LABOCA data) that constrains the fit to the steep-spectrum lobe emission, the subtraction of the non-thermal fit at 500\micron\ is highly uncertain and therefore not corrected or included in the Table. This is also true for the 350\micron\ fluxes of 1151-34, as can be seen below. Uncertainties in corrected fluxes  are those from the original \emph{Herschel} measurements, but do not account for uncertainties in the fitted power-law being subtracted. The original uncorrected fluxes are shown in brackets.  The core fits for PKS0915-11 and PKS2356-61 are based on the extrapolation of the fits to the LABOCA and 500\micron\ SPIRE data points. }
  \label{tab:NTcor}
  \begin{tabular}{@{}lllllllll@{}}
  \hline
   Name & 100$\umu$m & $\pm$  & 160$\umu$m & $\pm$   &  250$\umu$m & $\pm$    & 350$\umu$m & $\pm$ \\
 \hline
Steep Spectrum Fit &&&&&&&\\	
0023$-$26	&	49.0	(	58.5	)	&	4.3	&	72.0	(	86.6	)	&	8.6	&	49.6	(	71.7	)	&	5.0	&	34.9	(	65.0	)	&	6.9	\\
1151$-$34	&	41.1	(	47.8	)	&	2.0	&	40.7	(	51.2	)	&	3.8	&	15.8	(	31.8	)	&	5.8	&	-	(	24.2	)	&	-	\\
1814$-$63	&	147.0	(	186.3	)	&	5.2	&	133.2	(	189.0	)	&	7.6	&	57.2	(	135.1	)	&	7.7	&	28.2	(	128.3	)	&	9.4	\\
2135$-$20	&	75.0	(	80.8	)	&	5.0	&	126.0	(	134.3	)	&	6.8	&	79.0	(	91.3	)	&	4.4	&	35.0	(	55.0	)	&	6.4	\\
																													
Core fit &&&&&&&\\																													
0915$-$11	&	139.0	(	170.4	)	&	6.6	&	112.0	(	149.0	)	&	5.3	&	35.8	(	93.8	)	&	4.5	&	-				&	-	\\
2356$-$61	&	70.2	(	81.0	)	&	2.5	&	37.3	(	51.9	)	&	5.5	&	12.5	(	32.2	)	&	10.7	&	-				&	-	\\

\end{tabular}
 \end{table*}

\section{Discussion}
\label{sec:disc}

In this section we discuss the overall contribution of the AGN and associated jet activity to the far-IR continua of radio AGN, focussing first on non-thermal contamination and how it depends on 
the radio/optical classification, then on the possible contribution of AGN-heated cool dust.

\subsection{Dependence of non-thermal contamination on radio/optical class}
\label{sec:radio_optical}

Considering non-thermal contamination by both the lobes/hotspots and the cores of the radio sources, overall we find that a maximum of 43\% (PACS bands) and 72\% (SPIRE bands) of the objects in our full sample have significant non-thermal contamination, with the non-thermal contamination dominating the far-IR emission in at least 15\% (PACS bands) and 24\% (SPIRE bands) of objects. However, for 6 of the objects with non-thermal contamination in the PACS bands we can correct the contamination at 100 and 160\micron. Therefore, we are in a position to determine dust masses
-- one of the key goals of our project -- from the PACS photometry using the techniques of \citet{tadhunter14} for 70\% of the objects in our sample. This work is presented in \citet{bernhard22}.

Beyond assessing the incidence of non-thermal contamination in the sample as a whole, we find interesting differences in the rates of such contamination between the different radio AGN classifications, as highlighted by the results in Table \ref{tab:NTcontam}. We now discuss these differences in detail.

In terms of contamination by the extended lobe/hotspot emission, all the objects for which this contamination
is important are either compact CSS/GPS objects, or high redshift FRII sources. Indeed, 5/7 of the CSS/GPS sources in our sample
have substantial non-thermal lobe/hotspot emission at far-IR wavelengths, and thus lack evidence for the spectral steepening that might
otherwise indicate ageing of the electron populations in their radio lobes \citep[see][]{murgia99}. This is consistent with the young expansion ages for such sources derived from multi-epoch VLBI measurements \citep{owsianik98,murgia99,tschager2000,giroletti09,an12}. 

On the other hand, most of the objects with dominant or potential non-thermal core contamination are either broad-line radio galaxy or quasar FRII objects (BLRG/Q/FRII) or the weak-line radio galaxy FRI objects (WLRG/FRII).

Considering first the BLRG/Q/FRII, 5 out of 13 (39\%) of these objects are identified as having dominant or potential non-thermal core contamination at shorter far-IR wavelengths ($<$200\micron), and a further 3 objects show evidence for such contamination at longer far-IR wavelengths ($>$200\micron). Therefore, 8 out of 13 (61\%) of BLRG/FRII in the 2Jy sample are identified with clear or potential non-thermal core contamination to some part of their far-IR SED. In comparison, only 3 out of 15 (20\%) of the narrow-line radio galaxy Fanaroff-Riley Class II (NLRG/FRII) are identified with potential core contamination in the PACS bands, rising to 6 out of 15 (40\%) if we also consider the SPIRE bands. Although in terms of the rates of non-thermal contamination the difference between the BLRG/Q/FRII and NLRG/FRII is barely significant, it is notable that in four of the BLRG/Q/FRII the non-thermal component is clearly dominant in the SPIRE bands, whereas this is only the case for one of the NLRG/FRII objects.
Such differences are readily explained on the basis of the unified schemes for powerful radio galaxies: the jets of BLRG/Q/FRII objects are are likely to be inclined closer to our line of sight than for their NLRG/FRII counterparts (e.g. \citealp{barthel89,baldi13}), and therefore their far-IR SEDs are more likely to have a component of beamed emission from the inner radio jets. 

The two NLRG/FRII objects that show clearest evidence for non-thermal core contamination at far-IR wavelengths are PKS0442-28\footnote{Note that although its \emph{ALMA} and high frequency radio fluxes are high relative to the far-IR fluxes for this source, PKS0042-28 remains an uncertain case because of the lack of LABOCA and SPIRE flux measurements.} and PKS2356-61. It is unlikely that the prominence of the non-thermal core components in these cases is due to beaming effects, for example related to intermediate inclinations of their jets to the line of sight,
since the core-to-extended radio flux ratios of the two sources are typical of NLRG sources and
somewhat below the average values for BLRG/Q objects in the 2Jy sample \citep{morganti97}. Rather, it is more plausible that the relatively high level
of radio core contamination in these sources is related to their high overall radio luminosities: they represent the two objects with the highest total radio luminosities of the 20 FRII radio galaxies in the 2Jy sample at $z<0.15$ \citep{tadhunter98}.

Looking at the WLRG/FRI objects in Table \ref{tab:NTcontam}, it is notable that all 6 such objects have contamination from non-thermal core emission in their infrared SED. Indeed, the SEDs of 4 out of the 6 objects are completely dominated at far-IR wavelengths by the non-thermal core emission. This is consistent with previous results for FRI sources at both near- and mid-IR wavelengths which suggest a dominance of non-thermal components at these wavelengths, and an absence of warm dust emission from a circum-nuclear torus \citep{chiaberge99,leipski09,vanderwolk10}.

Of the two WLRG/FRI objects that are not dominated by non-thermal core emission at all far-IR wavelengths, PKS0915-11 (Hydra\,A) is particularly interesting: despite its FRI radio morphology, it is one of the most powerful radio sources in the local Universe. Also, unlike most WLRG/FRI sources, it has been identified to have a strong recent star formation component \citep{dicken12}. Therefore, it is possible that star formation heating of dust boosts the thermal component of its SED, so that the non-thermal emission does not dominate the infrared SED as for the other WLRG/FRI in the sample.  The other case -- PKS0625-53 -- is more ambiguous, since although its 160\micron\, flux falls below its 100\micron\, flux (possible evidence for a thermal bump), its 160\micron\, flux has relatively large uncertainties. Otherwise, the shape of the radio-to-IR SED of this object is consistent with it being dominated by non-thermal emission from the core, but with a fairly sharp cut-off at mid-IR wavelengths.

In stark contrast to the WLRG/FRI sources, the majority (4/5) of the WLRG/FRII sources  show clear evidence for thermal dust emission at far-IR wavelengths. The only exception is PKS0034-01, which is similar to the case of PKS0625-53 discussed above, with some uncertainty on whether or not it shows thermal dust emission, because of the relatively large error on its 160\micron\, flux\footnote{We note that PKS0034-01 is likely to have thermal dust emission at some level -- even if not clear from its far-IR SED -- given the detection of a dust lane in HST optical images \citep{martel99} and a star-forming disk in HST UV images \citep{baldi08}.}. The presence of a  cool ISM in these WLRG/FRII galaxies is consistent with the idea that they were originally SLRG/FRII, but that their AGN have recently switched off or entered a phase of reduced activity (see discussion in \citealp{tadhunter16}). However, far-IR observations of a larger sample of WLRG/FRII are required to put this idea on a firmer footing.

Finally, it is important to add the caveat that non-thermal emission from the cores of the sources is potentially subject to variability. Therefore, there may be some level of variability between the high frequency radio core measurements taken in 2006 with \emph{ATCA/VLA}, the \emph{Spitzer}/MIPS data from 2004 -- 2007,  the \emph{APEX} LABOCA data from 2011 -- 2013, the \emph{Herschel} data from 2012 -- 2013, and the \emph{ALMA} data from 2019. Such variability is hard to quantify for the objects in our sample. However, it may be apparent, for example, in the SED of PKS0625-35, which is dominated at all far-IR wavelengths by non-thermal core emission, but whose LABOCA and \emph{Herschel}  points -- well-represented by a single power-law -- fall above an interpolation of the \emph{ALMA} and \emph{Spitzer} MIPS points. This perhaps suggests that its non-thermal core emission was stronger in the period 2012 -- 2013 than it was either earlier or later. 

One clear sign of variability in the core would be if the \emph{ALMA} 100\,GHz flux were significantly higher than the \emph{ATCA/VLA} radio core flux: we would expect the core flux to decrease with increasing frequency, so a higher \emph{ALMA} flux may indicate variability. However, we only see such evidence in two objects: PKS0043-42 and PKS0442-28. 

Overall, in terms of contamination by the non-thermal cores, the presence of variability is unlikely to affect our general conclusions on the incidence of such contamination in the sample as a whole, since the highest frequency radio-to-mm flux we used for extrapolation purposes  could vary either upwards or downwards relative to the \emph{Herschel} fluxes. On the other hand, it could potentially affect our conclusions for some individual sources.

\subsection{Correlations between far-IR luminosities and AGN power indicators}
\label{sec:correlations}

As already mentioned in the introduction, the far-IR continuum luminosity is often considered to be a relatively clean indicator of the star formation activity in AGN host galaxies. Although attempts are sometimes made to correct the far-IR fluxes for AGN contamination using a template fitting approach \citep[e.g.][]{mullaney11,mor12,bernhard21}, many of the AGN templates used in such studies are made under the assumption that the AGN SED is steeply declining at far-IR wavelengths, as predicted by some models for the circum-nuclear torus surrounding the central energy-generating regions. However, \citet{symeonidis16} and \citet{symeonidis17} have   produced AGN SED templates that show a larger contribution to far-IR continuum, and have argued that much of the far-IR emission in luminous, quasar-like AGN is due to cool, AGN-heated dust.

Analysis of \emph{Spitzer} observations of the 2Jy sample also provides evidence for a major contribution from AGN-heated dust at far-IR wavelengths: in \citet{dicken09,dicken10} we found a strong correlation between the 70\micron\ continuum luminosity and the [OIII]$\lambda$5007 emission-line luminosity, which is often considered to provide a good indication of the AGN bolometric luminosity \citep[e.g.][]{heckman05}. Moreover, we showed that the AGN heating mechanism is energetically feasible if the cooler AGN-heated dust radiating at 70\micron\ has a covering factor similar to that of the narrow-line region (NLR). However, it could be argued that at 70\micron\ -- a relatively short far-IR wavelength -- the emission still has a significant contribution from dust in the circum-nuclear obscuring regions (e.g. the classical AGN torus), whereas at longer far-IR wavelengths the emission might be dominated by dust heated by regions of star formation, as a result of a steeply declining intrinsic AGN SED.

Our \emph{Herschel} observations allow us to further investigate the heating mechanism for the far-IR emitting dust, by extending the work of \citet{dicken09,dicken10} to longer far-IR wavelengths. We have used the flux measurements from Tables 1 and 3 to calculate the rest-frame far-IR continuum luminosities at 100 and 160\micron. As part of this, we made K corrections by assuming a modified black-body continuum shape with $\beta = 1.2$, and dust temperatures estimated using the 160/100\micron\ flux ratios \citep[see][for details]{tadhunter14}. For objects that were not detected at 160\micron, we assumed the mean cool dust temperature for all the objects free of non-thermal contamination and detected in both PACS bands to make the K-corrections. The derived 100 and 160\micron\ luminosities are presented in Table \ref{tab:luminosities} in the Appendix.

Note that we do not consider the 7 objects that are dominated by non-thermal emission in the PACS bands in this analysis, because we are interested in the relationship between the thermal far-IR emission and the AGN bolometric power. This removes all but two of the WLRG/FRI objects, so the analysis concentrates on the SLRG/FRII and WLRG/FRII objects. For the latter groups, we do not expect the removal of 3 non-thermally dominated objects to significantly bias the results, since on the basis of the unified schemes for radio-loud AGN these are objects whose jets are pointing close to the line of sight, but whose host galaxy properties are expected to be similar to those of the sample as a whole. We further note that, in cases where it has proved possible to correct the far-IR fluxes for non-thermal contamination (labelled with C in Table 2), we have used the corrected fluxes when calculating the luminosities.

\begin{figure} 
\begin{center}
\includegraphics [width=100mm]{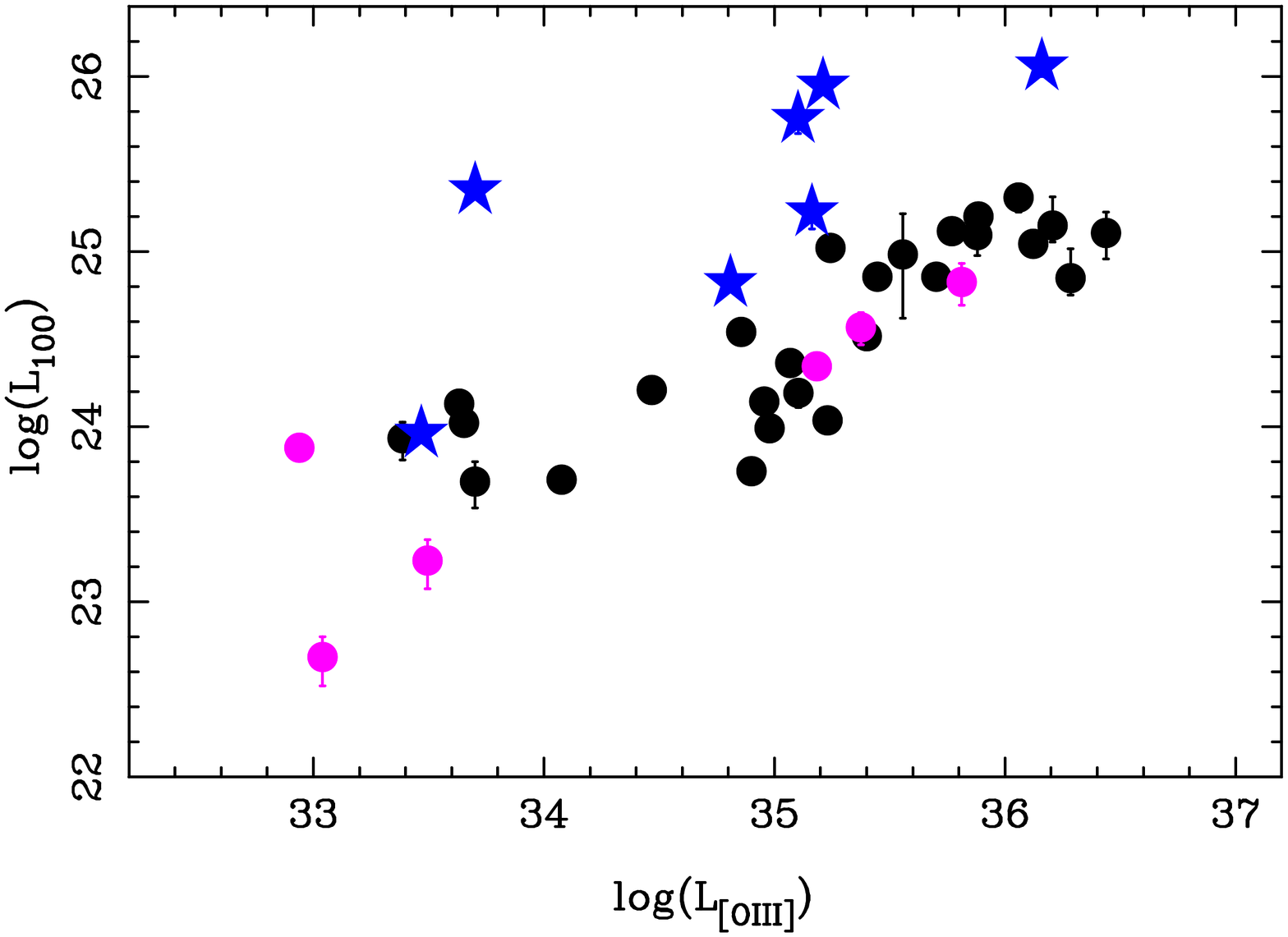}
\includegraphics [width=100mm]{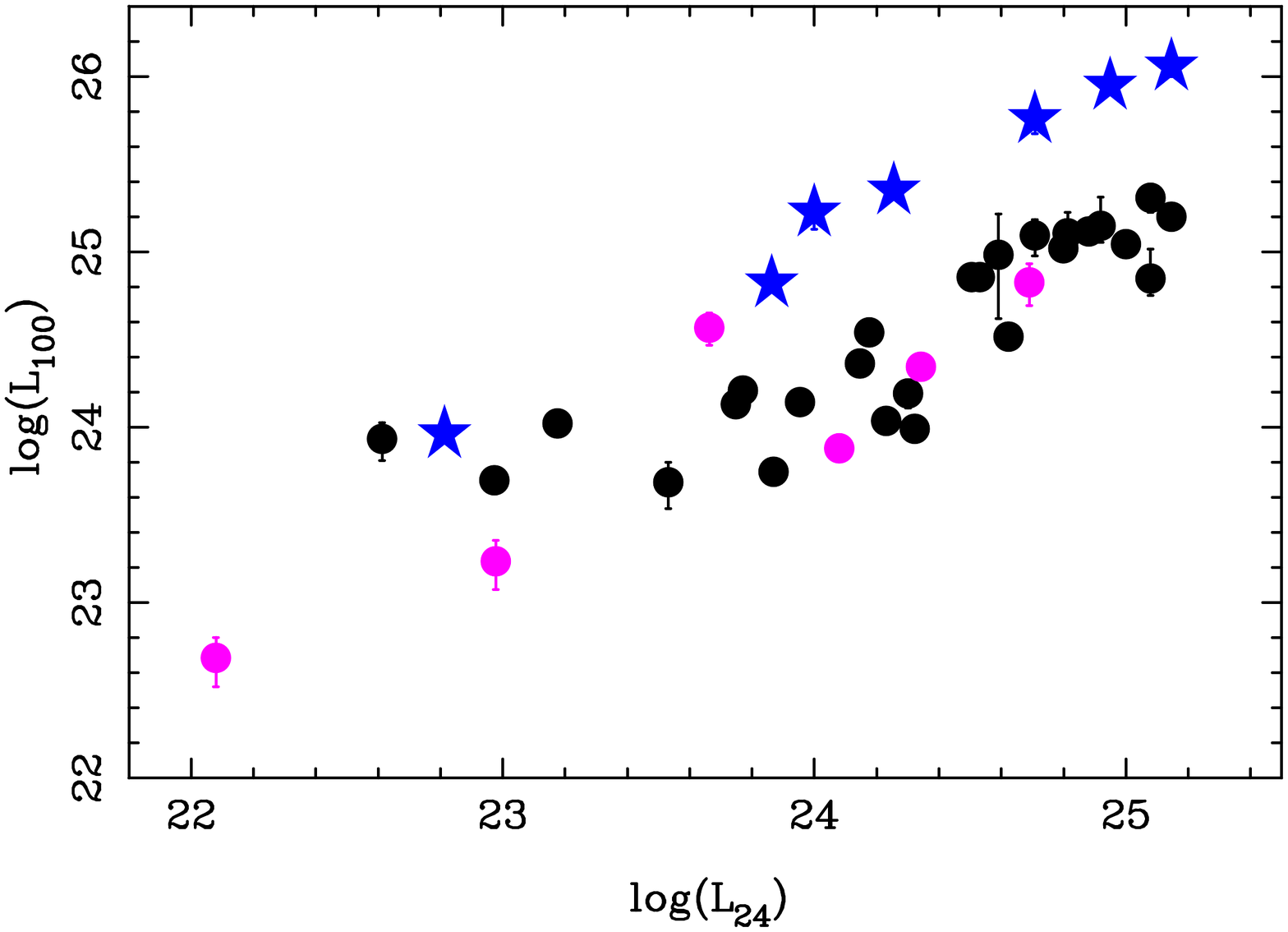}
\caption{The relationship between 100\micron\ far-IR luminosity and AGN bolometric power indicators. Top: $L_{100}$ vs. $L_{[OIII]}$. Bottom: $L_{100}$ vs. $L_{24}$. The 7 objects that are clearly dominated by non-thermal emission in the PACS bands are not plotted. Key to symbols: black circles indicate objects that do not suffer significant non-thermal contamination in the PACS bands or show independent evidence for strong star formation activity; magenta circles represent objects for which there is a possibility of non-thermal contamination in the PACS band; and blue stars are objects with clear evidence for recent star formation activity \citep[see][]{dicken12}. }
\label{Fig:Herschel_lum_100}
\end{center}
\end{figure}

\begin{figure} 
\begin{center}
\includegraphics [width=100mm]{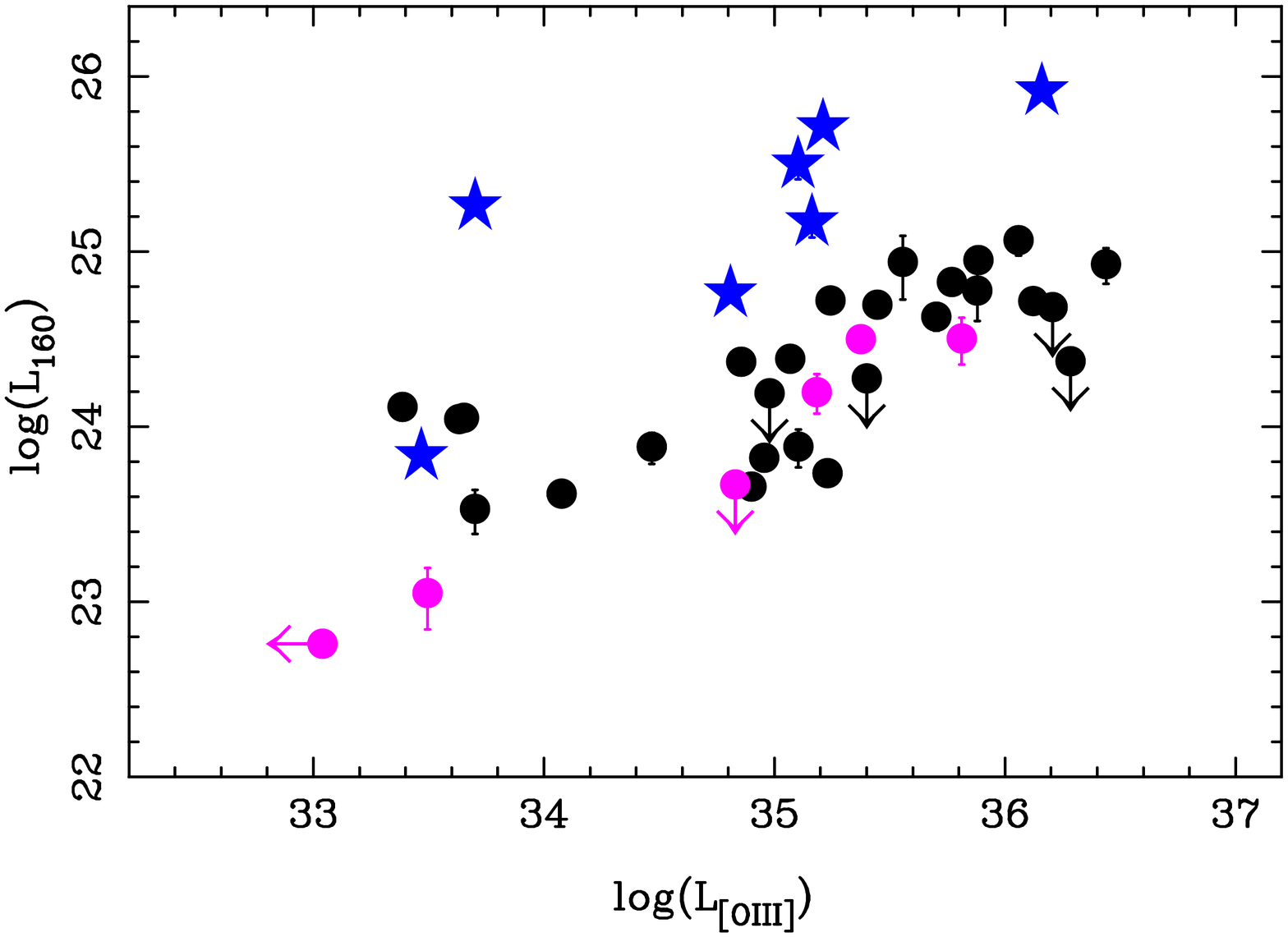}
\includegraphics [width=100mm]{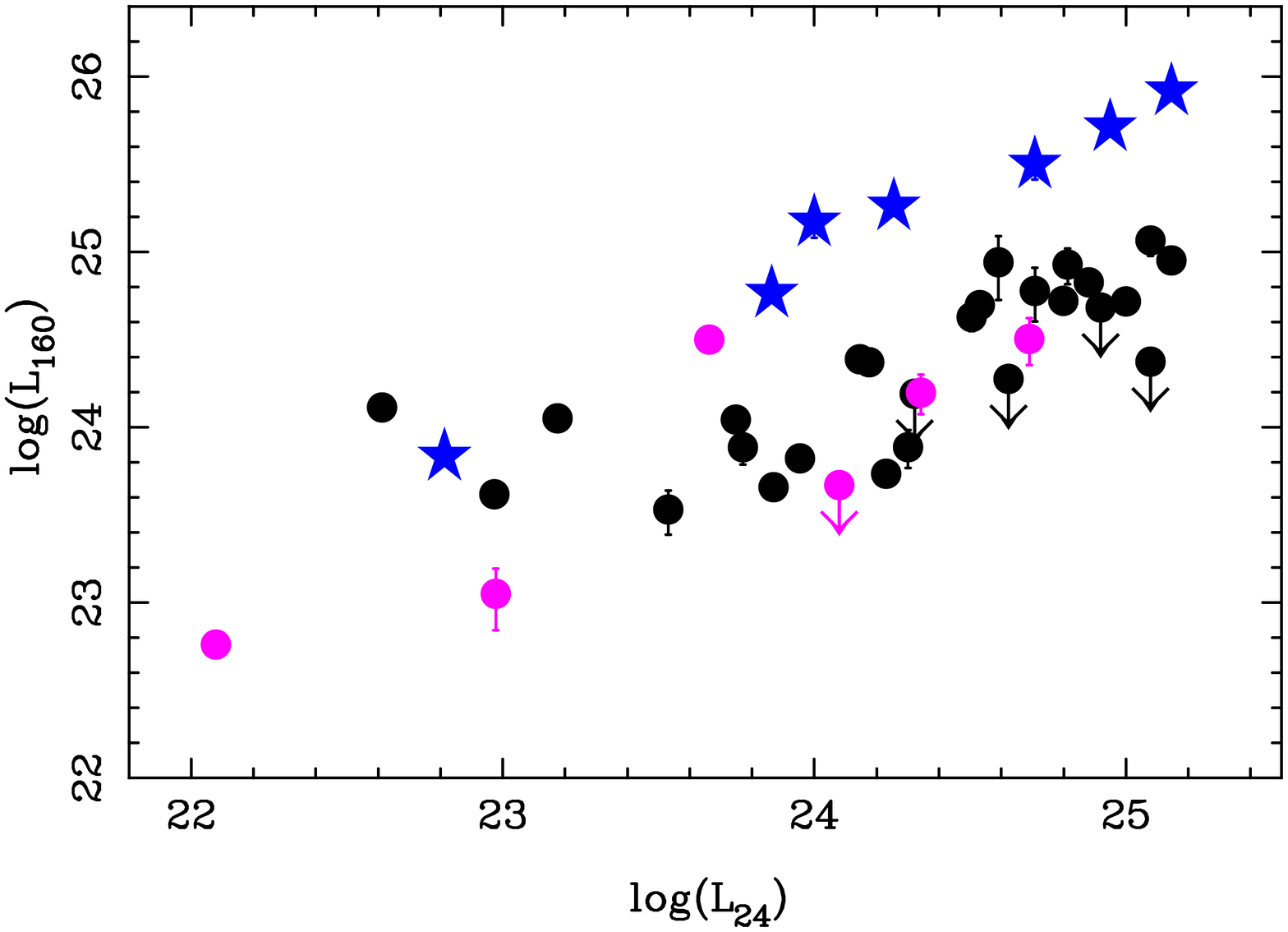}
\caption{The relationship between 160\micron\ far-IR luminosity and AGN bolometric power indicators. Top: $L_{160}$ vs. $L_{[OIII]}$. Bottom: $L_{160}$ vs. $L_{24}$. The 7 objects that are clearly dominated by non-thermal emission in the PACS bands are not plotted. The symbols are the same as in Figure \ref{Fig:Herschel_lum_100}.} 
\label{Fig:Herschel_lum_160}
\end{center}
\end{figure}

To investigate possible relationships between the far-IR luminosities and AGN bolometric power, we plot $L_{100}$ and $L_{160}$ against $L_{[OIII]}$ and $L_{24}$ in Figures \ref{Fig:Herschel_lum_100} and \ref{Fig:Herschel_lum_160}. Note that, like $L_{[OIII]}$,  $L_{24}$ is widely accepted as useful AGN bolometric indicator
\citep[e.g.][]{dicken14}\footnote{As shown in \citet{dicken14}, on average both $L_{[OIII]}$ and $L_{24}$ are likely to suffer a mild (factor $\sim$2) level of extinction in the type 2 compared with the type 1 objects in our sample. Such extinction effects will increase the scatter in the correlation plots, but not the overall result.}. At first sight, the plots in Figures \ref{Fig:Herschel_lum_100} and \ref{Fig:Herschel_lum_160} appear to provide evidence for correlations between the far-IR luminosities and the AGN bolometric luminosity indicators, with the correlations involving $L_{24}$ appearing to be tighter than those involving $L_{[OIII]}$. 

To put these correlations on a firm statistical footing, in the second column of Table \ref{tab:correlations} we present estimates of Spearman's rank correlation statistic ($\rho$)  for different combinations of luminosities, along with the probabilities that the luminosities are not correlated. For the correlations involving 160$\mu$m luminosities, we handled the small number of 160$\mu$m upper limits by taking the 100$\mu$m flux estimates (available for all objects), then for each affected object generating a 160$\mu$m flux using a 160/100$\mu$m ratio selected at random from the distribution of 160/100$\mu$m ratios measured for all the objects detected at both 160$\mu$m and 100$\mu$m. This process was repeated 100 times, with the correlation coefficients recalculated at each iteration. The final correlation coefficient values were then taken as the mean values for the 100 iterations.

Considering first the full sample of all 39 objects that are not dominated by non-thermal emission, the correlations are highly significant ($p < 0.5$\%). However, it is important to be cautious about spurious correlations that might arise through the mutual dependence of the luminosities on redshift, so we also present results for the Spearman partial rank correlation coefficient that take into account the potential redshift dependence (third column of Table \ref{tab:correlations}). On the basis of the latter, significant correlations are still found for $L_{100}$ vs $L_{24}$, $L_{100}$ vs $L_{[OIII]}$ and $L_{160}$ vs $L_{24}$, albeit with a reduced level of significance, but the $L_{160}$ vs $L_{[OIII]}$ correlation is no longer significant.

Before finally concluding that the far-IR luminosities are correlated with intrinsic AGN power, we need to consider other explanations for the trends in the plots of Figures \ref{Fig:Herschel_lum_100} and \ref{Fig:Herschel_lum_160}. One possibility is that the correlations involving $L_{24}$  -- which appear the tightest -- are  driven by non-thermal contamination: if the 24, 100 and 160$\mu$m luminosities were all affected by such contamination, then a correlation could arise without there being any correlation between the thermal dust emission and intrinsic AGN power. We have already removed all the objects dominated by non-thermal contamination (marked D in Table \ref{tab:NTcontam}) from the analysis, but now we also remove the objects with any possibility of non-thermal contamination in the PACS bands that cannot be corrected (marked P in Table \ref{tab:NTcontam}). The results are shown in the middle part of Table \ref{tab:correlations} and demonstrate that the removal of all the objects with any possibility of non-thermal contamination makes little difference to the correlation statistics.

\begin{table}
  \caption{Investigating of the correlations between $L_{\rm [OIII]}$, $L_{24}$, $L_{100}$ and $L_{160}$ using
  the Spearman rank correlation coefficient for different sub-samples.
  The
  final column gives the partial rank correlation statistics, which take into account any mutual
  dependence of the luminosities on redshift. The numbers in brackets give the percentage probabilities
  that the variables are uncorrelated.}
  \label{tab:correlations}
\begin{tabular}{lll}
\hline
Sample &Spearman's           &Spearman's \\
       &$\rho_{12}$\,(p\%)   &$\rho_{12,3}$\,(p\%) \\
\hline
Full sample, excluding & & \\
NT dominated ($N=39$) & & \\
$L_{\rm [OIII]}$ vs $L_{100}$ &0.711\,($<$0.5\%) &0.413\,($<$1\%) \\
$L_{\rm [OIII]}$ vs $L_{160}$ &0.646\,($<$0.5\%) &0.249\,($<$10\%) \\
$L_{24}$ vs $L_{100}$         &0.818\,($<$0.5\%) &0.623\,($<$0.5\%) \\
$L_{24}$ vs $L_{160}$         &0.747\,($<$0.5\%) &0.474\,($<$0.5\%) \\
\hline
Full sample, excluding NT & & \\ 
dominated+possible ($N=33$) & & \\ 
$L_{\rm [OIII]}$ vs $L_{100}$ &0.654\,($<$0.5\%) &0.376\,($<$2.5\%) \\
$L_{\rm [OIII]}$ vs $L_{160}$ &0.567\,($<$0.5\%) &0.180\,($>$10\%) \\
$L_{24}$ vs $L_{100}$         &0.793\,($<$0.5\%) &0.616\,($<$0.5\%) \\
$L_{24}$ vs $L_{160}$         &0.707\,($<$0.5\%) &0.427\,($<$1\%) \\
\hline
Full sample, excluding NT & & \\
dominated+possible and & & \\
strongly star forming ($N=26$) & & \\
$L_{\rm [OIII]}$ vs $L_{100}$ &0.849\,($<$0.5\%) &0.764\,($<$0.5\%) \\
$L_{\rm [OIII]}$ vs $L_{160}$ &0.659\,($<$0.5\%) &0.478\,($<$0.5\%) \\
$L_{24}$ vs $L_{100}$         &0.886\,($<$0.5\%) &0.823\,($<$0.5\%) \\
$L_{24}$ vs $L_{160}$         &0.818\,($<$0.5\%) &0.672\,($<$0.5\%) \\
\hline
\end{tabular}
 \end{table}

It is notable that the objects with clear, independent evidence for recent star formation activity (RSF) as indicated by  the detection of strong poly-aromatic hydrocarbon (PAH) features, optical/UV evidence for young stellar populations (UV excesses and/or detection of Balmer absorption lines), or red mid- to far-IR colours \citep{dicken12}, appear to show a particularly tight correlations in the plots involving $L_{24}$, and fall well above the main trends defined by the other objects in the sample. These RSF objects are indicated in the final column of Table \ref{tab:NTcontam}. For such objects, it is possible that the correlations could be driven by the fact that both the 24\micron\ and the far-IR luminosities are dominated by the thermal emission of dust heated by star formation, without there being a correlation between far-IR luminosity and intrinsic AGN power. To investigate the effect that this might have on the correlation statistics for the sample as a whole, we have removed the RSF objects, as well as those with any possibility of non-thermal contamination, and re-calculated the statistics. The results are presented in the bottom part of Table \ref{tab:correlations}, and show that all the correlation statistics, including the partial rank correlation statistics, suggest highly significant correlations between the 100 and 160\micron\ luminosities and intrinsic AGN power. It is interesting that we see the tightest correlations for this sub-sample, which is the cleanest in terms of potential non-thermal and strong star formation contamination at both mid- and far-IR wavelengths.

Overall, we find clear evidence for underlying correlation between far-IR luminosity and intrinsic AGN power. This is consistent with the results of \cite{dicken09,dicken10}, but now demonstrated for longer far-IR wavelengths. In terms of the underlying physical mechanisms at work, one possibility is that much of the cool, far-IR emitting dust is situated in the kpc-scale NLR and directly heated by the shorter wavelength radiation from the AGN \citep{tadhunter07,dicken09,dicken10} in a majority of objects. However, it has yet to be demonstrated that there is sufficient dust in the NLR on large enough scales, and hence cool enough to radiate in the far-IR, for this mechanism to be feasible. 

As well as the kpc-scale NLR, the circum-nuclear torus ($r <300$\,pc) may also make a significant contribution at far-IR wavelengths: the radiative transfer models of \citet{siebenmorgen15} show that, for certain assumed torus geometries and dust grain properties, the torus emission can fully account for far-IR SEDs of some AGN, without the need for a dust component heated by star formation. However, such models require a relatively large ratio of the outer to the inner torus radius ($r_{out}/r_{in} \sim 170$), as well as a significant population of large, ``fluffy'' dust grains with high far-IR emissivities that are not found in the normal ISM of galaxies. Again, since the dust in the torus is ultimately heated by shorter wavelength AGN radiation, a correlation might be expected between the far-IR luminosity and AGN bolometric luminosity, as
long as sufficient dust in the torus can remain cool enough to
radiate in the far-IR.

Regardless of whether the cool dust is situated in the NLR or the torus,  if it is mainly heated by the radiation of the AGN, this would have major implications for the use of far-IR continuum measurements to estimate SFR for the host galaxies of AGN. In particular, if the AGN heated cool dust component is not adequately accounted for,  estimates of SFR based on far-IR luminosities may be substantially over-estimated \citep[see also][]{symeonidis16,symeonidis17,symeonidis22}. 

On the other hand, our observations do not entirely rule out some level of heating of the cool dust by regions of star formation, even in cases that currently lack strong independent evidence for recent star formation activity. For example, the PAH features associated with low or moderate levels of star formation might be difficult to detect at mid-IR wavelengths in the face of strong, AGN-heated torus emission, and the optical/UV signs of star formation masked by AGN-related continuum components at shorter wavelengths \citep[e.g. scattered light and nebular continuum][]{tadhunter16}. Therefore, rather than direct heating of the cool dust by the AGN radiation, an alternative is that the observed trends arise indirectly via a correlation between AGN radiative power (heating the warmer mid-IR-emitting dust and ionizing the [OIII]-emitting regions) and star formation activity (heating the cooler far-IR-emitting dust). Such a correlation might be caused, for example, by the dependence of both the star formation rates and the accretion rates onto the central super-massive black holes on the supply of gas into the central regions of the host galaxies (e.g. via galaxy mergers).

\section{Conclusions}

We present deep \emph{Herschel} PACS and SPIRE observations for a complete sample of 46 2Jy southern radio AGN. The depths of the PACS observations have allowed $>$3$\sigma$ detections of 100 and 89\% of the full sample at 100 and 160\micron\ respectively. 19 objects were also observed with SPIRE, reaching a $>$3$\sigma$ detection rates of 84\% at 250\micron; 79\% at 350\micron; and 53\% at 500\micron.
In addition, to bridge the gap between far-IR and radio wavelengths, we present new sub-mm and mm data taken using \emph{APEX} LABOCA and \emph{ALMA} respectively, reporting 870\micron\ fluxes for 19 objects,  and $\sim$100\,GHz ($\sim$3\,mm) fluxes for 29 objects. 

We find that the infrared to radio SEDs of the sample show considerable diversity that is related to relative importance of the emission from the cores, jet and lobes of the radio sources, from the warm and cool dust heated by AGN radiation, and from the cool dust heated by regions of star formation. At far-IR wavelengths, the non-thermal emission is most prominent in CSS/GPS, BLRG/Q and WLRG/FRI sources. On the other hand, the thermal emission of cool dust heated by stars dominates in objects that show independent evidence for major recent star formation activity.

Despite the fact that all the objects in our sample were selected to have SEDs that are steeply declining with frequency at radio wavelengths, a significant fraction show evidence for non-thermal contamination in the far-IR, with the incidence of 
contamination higher in the longer-wavelength SPIRE bands (30 -- 72\%) than in the 
shorter wavelength PACS bands (28 -- 43\%). This emphasises that selection on the basis of a steep radio spectral index is not by itself sufficient to avoid non-thermal contamination in the far-IR. Moreover, even in cases where the extrapolation of steep-spectrum radio emission suggests negligible non-thermal contamination by the radio lobes, we have shown that contamination by the flatter-spectrum radio cores can be significant, or even dominate, at far-IR wavelengths.

Finally, concentrating on the subset of objects that lack independent evidence for strong star formation activity, and for which the non-thermal contamination in the PACS bands is negligible or can be corrected, we find highly significant correlations between
the 100 and 160\micron\ monochromatic continuum luminosities ($L_{100}$ and $L_{160}$) and AGN bolometric power indicators ($L_{[OIII]}$ and $L_{24}$). These correlations persist even after the potential mutual dependencies of the luminosities on redshift are taken into account. This provides further evidence that radiation from the AGN is an important heating source for the cool, far-IR emitting dust in many radio AGN, although we cannot entirely rule out the idea that the correlations arise through a strong relationship between star formation activity and AGN power.

\section*{Data Availability Statement}
The Herschel data underlying this article are available in the ESA Herschel Science Archive, at http://archives.esac.esa.int/hsa/whsa/. The ALMA data from this article are available in the ALMA Archive, at https://almascience.nrao.edu/alma-data. The APEX LABOCA data presented in this article can be shared on reasonable request to the authors.

\section*{Acknowledgments}

CT and EB acknowledge support from STFC.
We would like to thank Attila Kovacs for his help and support for the CRUSH software used for the \emph{APEX} LABOCA data reduction (https://sigmyne.com/crush/index.html). Based on observations made with ESO Telescopes \emph{APEX} Observatory (programs ID E-088.B-0947A-2011, E-089.B-0791A-2012, E-091.B-0217A-2013) and \emph{ALMA} (program IDs 2019.1.01022 and 2017.1.00629.S). \emph{Herschel} is an ESA space observatory with science instruments provided by European-led Principal Investigator consortia and with important participation from NASA. For the purpose of open access, the authors have applied a Creative Commons Attribution (CC BY) licence to any Author Accepted Manuscript version arising.

CRA acknowledges financial support from the Spanish Ministry of Science, Innovation
and Universities (MCIU) under grant with reference RYC-2014-15779, from
the European Union's Horizon
2020 research and innovation programme under Marie Sk\l odowska-Curie
grant agreement No 860744 (BiD4BESt), from the State
Research Agency (AEI-MCINN) and from the Spanish MCIU under grants
"Feeding and feedback in active galaxies" with reference
PID2019-106027GB-C42, "Quantifying the impact of quasar feedback on galaxy evolution (QSOFEED)", with reference EUR2020-112266, and from the Consejería de Econom\' ia, Conocimiento y Empleo del Gobierno de 
Canarias and the European Regional Development Fund (ERDF) under grant with reference ProID2020010105.




\bibliographystyle{mnras}
\bibliography{bib_list} 



\appendix
\setcounter{table}{0}
\renewcommand{\thetable}{A\arabic{table}}
\setcounter{figure}{0}
\renewcommand{\thefigure}{A\arabic{figure}}

\section{Notes on infrared observations for individual objects}

\begin{enumerate}
  \item {\bf PKS0235-19 -} The \emph{Spitzer} MIPS 70\micron\ flux first reported in \citet{dicken08} now appears to be high compared with the new \emph{Herschel} photometry results. Re-analysing the \emph{Spitzer} MIPS data, what previously appeared to be the object may be a feature in the background, where we find the \emph{Spitzer} image to be particularly poor in quality. 
  \item {\bf PKS0404+03 -} \emph{Herschel} SPIRE data flux measurements are not possible because of bright Galactic cirrus emission in the images. 
    \item {\bf PKS0620-52 -} Measurement of the \emph{Herschel} PACS 160\micron\ flux was difficult due to two close neighbours in the image for this object. The flux reported is the sum of all 3 objects fluxes minus the two brighter companion galaxies. Note that the  70\micron\ flux for this object is unreliable because of the potential contamination by the companion objects in the \emph{Spitzer} beam.
  
  \item {\bf PKS0945+07 -} The \emph{Spitzer} MIPS 70\micron\ flux first reported in \citet{dicken08} now appears to be low compared with the new \emph{Herschel} photometry results. Re-examination the \emph{Spitzer} MIPS data reveals negative pixels in the vicinity of the source that may have brought down the overall flux in this object. 
   \item {\bf PKS0915-11 -} At 350 and 500\micron\ this source is clearly confused with its neighbour -- a spiral galaxy $\sim$35\,arcsec to the SW \citep{ramos11a} -- in the SPIRE images. 
   The grey points plotted in the SED of PKS0915-11 are the PACS fluxes for the companion object that is confused with the target at the two longer SPIRE bands. These points are plotted to help assess the degree of potential contamination to the longer SPIRE bands in this object. 
  \item {\bf PKS1306-09  -} The \emph{Spitzer} MIPS 70\micron\ flux first reported in \citet{dicken08} now appears to be high compared with the new \emph{Herschel} photometry results. The \emph{Herschel} PACS measurements fall on a straightforward power-law extrapolation of the radio to sub-mm SED, but the \emph{Spitzer} MIPS 70\micron\ point falls significantly above this line. 
   \item {\bf PKS1934-63  -}  The PACS 100 and 160\micron\ fluxes have been corrected for the contribution of a companion galaxy ~17 arcsec to the SW; however, it was not possible to correct the \emph{Spitzer} 70\micron\ flux for the companion, so this may represent an upper limit for the PKS1934-63 AGN host.
   \item {\bf PKS2135-14  -}   The \emph{Spitzer} MIPS 70\micron\ flux first reported in \citet{dicken08} now appears to be low compared with the new \emph{Herschel} photometry results. Re-analysing the \emph{Spitzer} MIPS data reveals negative pixels in the vicinity of the source that may have brought down the overall flux in this object. 
   \item {\bf PKS2250-41  -}   The \emph{Spitzer} MIPS 70\micron\ flux first reported in \citet{dicken08} now appears to be low compared with the new \emph{Herschel} photometry results. 
   \item {\bf PKS2221-02  -}   The flux drops rapidly for this object at SPIRE wavelengths, and the SPIRE images are likely to be dominated by the flux of a nearby companion galaxy that is clearly visible in the PACS bands (companion fluxes shown by grey triangles in the SED of Fig 7).

\end{enumerate}

 \begin{table*}
  \caption{Rest-frame [OIII]$\lambda$5007, 5GHz radio, 24\micron, 100\micron\ and 160\micron\ luminosities for the 2Jy sample. The [OIII], radio and 24$\mu$m luminosities are taken from \citet{dicken09}, with the exception of PKS0347+05, whose [OIII] luminosity has been revised based on the work of \citet{tadhunter12}; the 100 and 160$\mu$m luminosities are derived from the \emph{Herschel} fluxes presented in this paper (see
Section 2.2 for details). Note that the K-corrections used to calculate the far-IR luminosities of  objects not detected at 160\micron\, or dominated by non-thermal radiation, were derived by assuming the mean dust temperature ($T_{dust} = 40.1$\,K), as determined for all the objects without significant star formation or evidence for non-thermal contamination. The far-infrared luminosities of the non-thermal dominated objects -- indicated by brackets in the table -- are uncertain, because is not clear that the assumed modified black body is a good approximation to their true SED shapes.  }

  \label{tab:luminosities}
  \begin{center}
  \begin{tabular}{lccccc}
  \hline
   PKS     &$L_{[OIII}$  &$L_{5GHz}$  &$L_{24}$ &$L_{100}$ &$L_{160}$ \\
   Name &log(W) &log(W Hz$^{-1}$) &log(W Hz$^{-1}$) &log(W Hz$^{-1}$) &log(W Hz$^{-1}$) \\
 \hline
0023$-$26	&35.18 &27.04 &24.00 &25.22 &25.17 \\
0034$-$01	&33.49 &25.30 &22.98 &23.23 &23.05 \\
0035$-$02	&35.08 &26.23 &24.20 &(24.59) &(24.52) \\
0038$+$09	&35.17 &26.20 &24.34 &24.34 &24.20		\\
0039$-$44	&36.06 &26.66 &25.08 &25.31 &25.06  \\
0043$-$42	&33.70 &25.97 &23.53 &23.69 &23.53 \\
0105$-$16	&35.39 &26.83 &24.62 &24.52 &$<$24.28 \\
0117$-$15	&36.21 &27.23 &24.91 &25.15 &$<$24.68 \\
0213$-$13	&35.10 &26.00 &24.30 &24.19 &23.89	\\
0235$-$19	&36.27 &27.34 &25.08 &24.85 &$<$24.38 \\
0252$-$71	&35.15 &27.34 &24.59 &24.98 &24.94	\\
0347$+$05	&33.70 &26.65 &24.26 &25.35 &25.26		\\
0349$-$27	&34.08 &25.32 &22.97 &23.70 &23.61	\\
0404$+$03	&34.46 &26.65 &23.77 &24.21 &23.89 \\
0409$-$75	&35.11 &27.91 &24.71 &25.76 &25.50	\\
0442$-$28	&34.83 &26.08 &24.08 &23.88 &$<$23.67	\\
0620$-$52	&$<$32.41 &24.87 &22.46 &(22.99) &(22.98)	\\
0625$-$35	&33.48 &25.17 &23.23 &(23.70) &(23.74)	\\
0625$-$53	&$<$33.04 &25.04 &22.08 &22.69 &22.76 \\
0806$-$10	&35.77 &25.65 &24.88 &25.12 &24.83	\\
0859$-$25	&34.98 &26.72 &24.32 &23.99 &$<$24.19		\\
0915$-$11	&33.46 &25.96 &22.81 &25.96 &23.83	\\
0945$+$07	&34.90 &25.67 &23.87 &23.75 &23.66		\\
1136$-$13	&36.73 &27.30 &25.14 &(25.44) &(25.13)		\\
1151$-$34	&35.45 &26.72 &24.53 &24.86 &24.70	\\
1306$-$09	&35.15 &27.11 &24.64 &(25.11) &(25.00)	\\
1355$-$41	&35.89 &26.65 &25.15 &25.20 & 24.95\\
1547$-$79	&36.43 &27.08 &24.81 &25.11 &24.93		\\
1559$+$02	&35.25 &25.89 &24.80 &25.02 &24.72	\\
1602$+$01	&35.81 &26.89 &24.69 &24.83 &24.50		\\
1648$+$05	&33.65 &26.91 &23.18 &24.02 &24.05	\\
1733$-$56	&34.81 &25.89 &23.86 &24.82 &24.76	\\
1814$-$63	&33.63 &25.52 &23.75 &24.13 &24.04	\\
1839$-$48	&$<$32.36 &25.60 &23.00 &(23.52) &(23.96)	\\
1932$-$46	&35.38 &26.74 &23.66 &24.57 &24.50	\\
1934$-$63	&35.08 &26.77 &24.15 &24.36 &24.39	\\
1938$-$15	&35.88 &27.32 &24.70 &25.09 &24.78	\\
1949$+$02	&34.86 &25.27 &24.17 &24.54 &24.37	\\
1954$-$55	&$<$32.0 &25.15 &22.36 &(22.88) &(22.90) \\
2135$-$14	&36.11 &26.18 &25.00 &25.04 &24.71	\\
2135$-$20	&36.14 &27.38 &25.14 &26.06 &25.91	\\
2211$-$17	&33.38 &26.15 &22.61 &23.93 &24.11	\\
2221$-$02	&35.23 &25.23 &24.23 &24.04 &23.73	\\
2250$-$41	&35.70 &25.59 &24.51 &24.86 &24.63	\\
2314$+$03	&35.20 &26.25 &24.94 &25.95 &25.71 \\
2356$-$61	&34.95 &26.00 &23.95 &24.14 &23.82 	\\

\end{tabular}
\end{center}
 \end{table*}

 \begin{table*}
 \begin{minipage}{175mm}
  \caption{{Observation details for the \emph{Herschel}, \emph{APEX/LABOCA} and \emph{ALMA} data. }}
  \label{tab:obs_param}
  \begin{tabular}{@{}l ll l l l l l c l r @{}} 

  \hline
	&	\emph{Herschel}  &		&	&	 &	&	&	\emph{APEX}	&		&\emph{ALMA}	&		 \\
  \hline	
	&	PACS &	&	&	 &SPIRE	&	&	LABOCA	&		&	&		 \\
  \hline
   PKS 	& Obs. &	Obs.  	&	Duration   &Obs. &	Obs. 	&	duration 	&	Obs. 	&	Obs.  		&Obs. 	&	LO Freq 	  \\
   
    Name	&IDs &	Date 	&	blue/red (s)	&ID &	 Date  &	 (s)	&	Date 	&	 prog. 		&Date 	&	(GHz)		 \\
   \hline
0023$-$26 &1342257778/79	 &	25/12/12	&	220	 &1342258392	&	3/01/13	&	721	&		&		&		&				\\
0034$-$01 &1342257780/81	&	26/12/12	&	670	&	&		&		&		&		&	10/11/19	&	102.4		\\
0035$-$02 &1342258794/95	&	6/01/13	&	220	&1342258370	&	3/01/13	&	721	&	15/09/11	&	88	&	3/11/19	&	101.4		\\
0038$+$09 &1342258800/01	&	6/01/13	&	670	&	&		&		&		&		&	1/11/19	&	91.9		\\
0039$-$44 &1342259082/83	&	9/01/13	&	220	&1342258401	&	3/01/13	&	721	&	30/04/13	&	91	&		&				\\
0043$-$42 &1342256973/74	&	11/1/12	&	1570	&	&		&		&		&		&	2/11/19	&	98.2		\\
0105$-$16 &1342258812/13	&	6/01/13	&	1570	&	&		&		&		&		&		&				\\
0117$-$15 &1342258810/11	&	6/01/13	&	670	&	&		&		&		&		&		&				\\
0213$-$13 &1342258808/09	&	6/01/13	&	670	&	&		&		&		&		&	3/11/19	&	95.4		\\
0235$-$19 &1342259276/77	&	10/01/13	&	1570	&	&		&		&		&		&		&				\\
0252$-$71 &1342253004/05	&	10/10/12	&	1570	&	&		&		&	1/08/12	&	89	&		&				\\
0347$+$05 &1342263912/13	&	19/02/13	&	670	&	&		&		&		&		&		&				\\
0349$-$27 &1342261724/25	&	21/01/13	&	670	&	&		&		&	14/04/13	&	91	&	4/11/19	&	103.1		\\
0404$+$03 &1342267196/97	&	12/03/13	&	670	&1342265381	&	2/03/13	&	721	&		&		&	3/11/19	&	100.9		\\
0409$-$75 &1342252998/99	&	10/10/12	&	670	&	&		&		&	1/08/12	&	89	&		&				\\
0442$-$28 &1342263893/94	&	18/02/13	&	670	&	&		&		&		&		&	3/11/19	&	95.4		\\
0620$-$52 &1342253013/14	&	10/10/12	&	670	&	&		&		&	1/08/12	&	89	&	3/11/19	&	104.6		\\
0625$-$35 &1342252855/56	&	9/10/12	&	220	&1342253431	&	15/10/12	&	721	&	11/04/13	&	91	&	10/11/19	&	104.3		\\
0625$-$53 &1342253011/12	&	10/10/12	&	1570	&		&		&		&		&		&	3/11/19	&	104.3		\\
0806$-$10 &1342253031/32	&	10/10/12	&	220	&1342254513	&	5/11/12	&	721	&		&		&	4/11/19	&	98.9		\\
0859$-$25 &1342253027/28	&	10/10/12	&	1570	&		&		&		&		&		&		&				\\
0915$-$11 &1342207071/2/3/4	&	25/10/10	&	571	&1342207041	&	24/10/10	&   721	& 21/04/13	&	91	& 18/07/18		& 102.6				\\
0945$+$07 &1342255960/61	&	27/11/12	&	1570	&	&		&		&		&		&	3/11/19	&	101.1		\\
1136$-$13 &1342257457/58	&	19/12/12	&	1570	&		&		&		&	13/04/13	&	91	&		&				\\
1151$-$34 &1342257565/66	&	20/12/12	&	670	&1342259409	&	3/01/13	&	721	&	13/04/13	&	91	&	3/11/19	&	98.6		\\
1306$-$09 &1342257563/64	&	20/12/12	&	670	&	&		&		&	22/05/12	&	89	&		&				\\
1355$-$41 &1342259286/87	&	10/01/13	&	445	&1342261665	&	21/01/13	&	721	&		&		&		&				\\
1547$-$79 &1342252992/93	&	10/10/12	&	670	&	&		&		&		&		&		&				\\
1559$+$02 &1342261307/08	&	17/01/13	&	220	&1342261672	&	21/01/13	&	721	&		&		&	14/11/19	&	99.4		\\
1602$+$01 &1342261305/06	&	17/01/13	&	1570	&	&		&		&	13/04/13	&	91	&		&				\\
1648$+$05  &1342252009/10   	&	01/10/12	&	1570	&1342251952	&	01/10/12	&	721	&		&		&	8/11/19	&	94.88		\\
1733$-$56 &1342252845/46	&	9/10/12	&	220	&1342253432	&	15/10/12	&	721	&	19/09/11	&	88	&	14/11/19	&	99.9		\\
1814$-$63 &1342252847/48	&	9/10/12	&	220	&1342254066	&	30/10/12	&	721	&	13/04/13	&	91	&	12/11/19	&	103.3		\\
1839$-$48 &1342252839/40	&	9/10/12	&	1570	&	&		&		&	13/04/13	&	91	&	30/10/19	&	98.6		\\
1932$-$46 &1342252837/38	&	9/10/12	&	1570	&	&		&		&	22/05/12	&	89	&	30/10/19	&	100.4		\\
1934$-$63 &1342252988/89	&	10/10/12	&	1570	&	&		&		&		&		&	30/10/19	&	92.2		\\
1938$-$15 &1342252984/85	&	10/10/12	&	1570	&	&		&		&		&		&		&				\\
1949$+$02 &1342252980/81	&	10/10/12	&	220	&1342255085	&	15/11/12	&	721	&		&		&	2/11/19	&	103.8			\\
1954$-$55 &1342252986/87	&	10/10/12	&	1570	&	&		&		&		&		&	2/11/19	&	103.9		\\
2135$-$14 &1342256130/31	&	27/11/12	&	220	&1342256742	&	8/12/12	&	721	&	22/05/12	&	89	&	31/10/19	&	102.9		\\
2135$-$20 &1342255980/81	&	27/11/12	&	220	&1342256740	&	8/12/12	&	721	&	22/05/12	&	89	&		&				\\
2211$-$17 &1342257509/10	&	19/12/12	&	1570	&	&		&		&		&		&	31/10/19	&	94.9		\\
2221$-$02 &1342257621/22	&	20/12/12	&	220	&1342256798	&	8/12/12	&	721	&		&		&	3/11/19	&	104.0		\\
2250$-$41 &1342256140/41	&	27/11/12	&	670	&	&		&		&		&		&		&				\\
2314$+$03 &1342237979/80	&	06/01/12	&	445	&1342234756	&	19/12/11	&	307	&		&		&	31/10/19	&	101.3		\\
2356$-$61 &1342257001/02	&	12/12/12	&	670	&1342259397	&	3/01/13	&	721	&	27/04/13	&	91	&	31/10/19	&	100.1		\\

\end{tabular}
\end{minipage}
 \end{table*}


\bsp	
\label{lastpage}
\end{document}